\documentclass[journal,twoside]{IEEEtran}
\IEEEoverridecommandlockouts
\usepackage{tikz}
\usepackage{pgfplots}
\usepackage{bm}
\usepackage{amsmath} 
\usepackage{amsthm}
\usepackage{amsfonts}
\usepackage{graphicx}
\usepackage{siunitx}
\usepackage{color}
\usepackage{fancyhdr}
\usepackage{array,multirow}
\usepackage{adjustbox}
\usepackage{balance}
\usepackage{cite}
\usepackage[acronym,shortcuts]{glossaries}
\usepackage{xspace}
\usepackage{tumcolor}
\usepackage{hyperref}
\usetikzlibrary{calc, shapes.geometric, arrows, positioning, calc}

\usepackage{cleveref}

\pgfplotsset{
	discard if not/.style 2 args={
		x filter/.code={
			\edef\tempa{\thisrow{#1}}
			\edef\tempb{#2}
			\ifx\tempa\tempb
			\else
			\fi
		}
	}
}
\pgfplotsset{compat=1.15}
\definecolor{mylila}{RGB}{153,50,204} 
\definecolor{mygreen}{RGB}{176,191,26} 

\def\pltw{245pt}
\def\plth{140pt}
\def\plthsmall{110pt}
\def\markSize{1.8pt}
\def\lineWidth{1.0pt}

\tikzset{VAEgenie/.style={mark options=solid, color=TUMBeamerRed, line width=\lineWidth, mark=Mercedes star, mark size=\markSize, solid}}
\tikzset{VAEnoisy/.style={mark options={solid}, color=TUMBeamerOrange, line width=\lineWidth, mark=triangle, mark size=\markSize, solid}}
\tikzset{VAEreal/.style={mark options={solid}, color=mygreen, line width=\lineWidth, mark=pentagon, mark size=\markSize, solid}}
\tikzset{geniecov/.style={mark options={solid}, color=blue, line width=\lineWidth, mark=x, mark size=\markSize, dashed}}
\tikzset{globalcov/.style={mark options={solid}, color=TUMBeamerGreen, line width=\lineWidth, mark=|, mark size=\markSize, dashed}}
\tikzset{LS/.style={mark options={solid}, color=black, line width=\lineWidth, mark=Mercedes star flipped, mark size=\markSize, dashed}}
\tikzset{GMM/.style={mark options={solid}, color=brown, line width=\lineWidth, mark=o, mark size=\markSize, dashed}}
\tikzset{{GMM circ}/.style={mark options={solid}, color=mylila, line width=\lineWidth, mark=diamond, mark size=\markSize, dashed}}
\tikzset{{GMM kron}/.style={mark options={solid}, color=brown, line width=\lineWidth, mark=o, mark size=\markSize, dashed}}
\tikzset{{GMM bcirc}/.style={mark options={solid}, color=mylila, line width=\lineWidth, mark=diamond, mark size=\markSize, dashed}}
\tikzset{CNN/.style={mark options={solid}, color=TUMBeamerLightBlue, line width=\lineWidth, mark=10-pointed star, mark size=\markSize, dashed}}
\tikzset{AMP/.style={mark options={solid}, color=TUMBeamerDarkRed, line width=\lineWidth, mark=star, mark size=\markSize, dashed}}

\newcommand{\legVAEgenie}{VAE-genie}
\newcommand{\legVAEnoisy}{VAE-noisy}
\newcommand{\legVAEreal}{VAE-real}
\newcommand{\leggeniecov}{genie-cov}
\newcommand{\legglobalcov}{global-cov}
\newcommand{\legLS}{LS}

\newcommand{\leggmmcirc}{GMM}

\newcommand{\leggmmbcirc}{GMM}
\newcommand{\legcnn}{CNN}
\newcommand{\legamp}{AMP}

\def\arrowscale{1.25}

\newcommand{\quadriga}{QuaDRiGa\xspace}
\newcommand{\mapvae}{\ac{map}-\ac{vae}\xspace}
\newcommand{\Ntx}{{N_\text{tx}}}
\newcommand{\Nrx}{{N_\text{rx}}}
\newcommand{\Np}{{N_\text{p}}}
\newcommand{\NL}{{N_\text{L}}}
\newcommand{\cnd}{{\,|\,}}

\newcommand{\herm}{{^{\mkern0.5mu\mathrm{H}}}}
\newcommand{\tran}{{^{\mkern0.5mu\mathrm{T}}}}

\newcommand{\msig}{{\bm{\Sigma}}}
\newcommand{\vdel}{{\bm{\delta}}}
\newcommand{\vmu}{{\bm{\mu}}}
\newcommand{\vsig}{{\bm{\sigma}}}
\newcommand{\veps}{{\bm{\varepsilon}}}
\newcommand{\vtheta}{{\bm{\theta}}}
\newcommand{\vphi}{{\bm{\phi}}}

\newcommand{\va}{{\bm{a}}}
\newcommand{\vb}{{\bm{b}}}
\newcommand{\vc}{{\bm{c}}}

\newcommand{\vh}{{\bm{h}}}

\newcommand{\vn}{{\bm{n}}}

\newcommand{\vy}{{\bm{y}}}
\newcommand{\vz}{{\bm{z}}}

\newcommand{\ma}{{\bm{A}}}
\newcommand{\mb}{{\bm{B}}}
\newcommand{\mc}{{\bm{C}}}
\newcommand{\md}{{\bm{D}}}

\newcommand{\mf}{{\bm{F}}}
\newcommand{\mg}{{\bm{G}}}
\newcommand{\mh}{{\bm{H}}}

\newcommand{\mn}{{\bm{N}}}

\newcommand{\mq}{{\bm{Q}}}

\newcommand{\mx}{{\bm{X}}}
\newcommand{\my}{{\bm{Y}}}

\newcommand{\RR}{{\mathbb{R}}}
\newcommand{\CC}{{\mathbb{C}}}
\newcommand{\diff}{{\mathrm{d}}}
\newcommand{\jim}{{\mathrm{j}}}
\newcommand{\OO}{{\mathcal O}}

\DeclareMathOperator{\E}{E}
\DeclareMathOperator{\vect}{vec}
\DeclareMathOperator{\diag}{diag}
\DeclareMathOperator{\KL}{D_{KL}}
\DeclareMathOperator*{\argmin}{arg\,min}

\newcommand{\tr}{\operatorname{tr}}
\newcommand{\B}[1]{{\bm{#1}}}
\newcommand{\btheta}{{\B \theta}}
\newcommand{\bphi}{{\B \phi}}
\newcommand{\Et}{\E_{\B z}}
\newcommand{\fto}{f_{\btheta,1}(\B z)}
\newcommand{\ftt}{f_{\btheta,2}(\B z)}
\newcommand{\fmo}{f_{\btheta,1}(\B \mu_{\bphi})}
\newcommand{\fmt}{f_{\btheta,2}(\B \mu_{\bphi})}
\DeclareMathOperator{\eye}{\mathbf{I}}
\newcommand{\inv}{^{-1}}

\newacronym{tdd}{TDD}{time division duplex}
\newacronym{fdd}{FDD}{frequency division duplex}
\newacronym{lmmse}{LMMSE}{linear minimum mean squared error}
\newacronym{mse}{MSE}{mean squared error}
\newacronym{mmse}{MMSE}{minimum mean squared error}
\newacronym{nmse}{NMSE}{normalized mean squared error}
\newacronym{mimo}{MIMO}{multiple-input multiple-output}
\newacronym{simo}{SIMO}{single-input multiple-output}
\newacronym{miso}{MISO}{multiple-input single-output}
\newacronym{siso}{SISO}{single-input single-output}
\newacronym{deep}{DL}{deep learning}
\newacronym{ofdm}{OFDM}{orthogonal frequency division multiplexing}
\newacronym{csi}{CSI}{channel state information}
\newacronym{ula}{ULA}{uniform linear array}
\newacronym{ura}{URA}{uniform rectangular array}
\newacronym{dft}{DFT}{discrete fourier transform}
\newacronym{bs}{BS}{base station}
\newacronym{mt}{MT}{mobile terminal}
\newacronym{ae}{AE}{autoencoder}
\newacronym{ml}{ML}{machine learning}
\newacronym{dl}{DL}{deep learning}
\newacronym{doa}{DoA}{direction of arrival}
\newacronym{dod}{DoD}{direction of departure}
\newacronym{kl}{KL}{Kullback-Leibler}
\newacronym{elbo}{ELBO}{evidence lower bound}
\newacronym{iid}{i.i.d.}{independent and identically distributed}
\newacronym{fc}{FC}{fully connected}
\newacronym{nn}{NN}{neural network}
\newacronym{dnn}{NN}{neural network}
\newacronym{cnn}{CNN}{convolutional neural network}
\newacronym{ls}{LS}{least squares}
\newacronym{snr}{SNR}{signal-to-noise ratio}
\newacronym{ce}{CE}{channel estimation}
\newacronym{ul}{UL}{uplink}
\newacronym{mpc}{MPC}{multipath component}
\newacronym{rt}{RT}{ray-tracing}
\newacronym{mmd}{MMD}{maximum mean discrepancy}
\newacronym{cdf}{CDF}{cumulative distribution function}
\newacronym{pdf}{PDF}{probability density function}
\newacronym{tpr}{TPR}{true positive rate}
\newacronym{ccm}{CCM}{channel covariance matrix}
\newacronym{cg}{CG}{conditionally Gaussian}
\newacronym{3gpp}{3GPP}{3rd Generation Partnership Project}
\newacronym{vae}{VAE}{variational autoencoder}
\newacronym{vi}{VI}{variational inference}
\newacronym{cc}{CC}{convolutional channel}
\newacronym{cl}{CL}{convolutional layer}
\newacronym{ll}{LL}{linear layer}
\newacronym{rl}{RL}{reshaping layer}
\newacronym{bn}{BN}{batch normalization}
\newacronym{cs}{CS}{compressed sensing}
\newacronym{amp}{AMP}{approximate message passing}
\newacronym{gmm}{GMM}{Gaussian mixture model}
\newacronym{nf}{NF}{normalizing flow}
\newacronym{gan}{GAN}{generative adversarial network}
\newacronym{dm}{DM}{diffusion model}
\newacronym{vdm}{VDM}{variational diffusion model}
\newacronym{cme}{CME}{conditional mean estimator}
\newacronym{los}{LOS}{line of sight}
\newacronym{nlos}{NLOS}{non-line of sight}
\newacronym{rv}{RV}{random variable}
\newacronym{gm}{GM}{generative model}
\newacronym{psd}{PSD}{positive semi-definite}
\newacronym{wss}{WSS}{wide-sense stationary}
\newacronym{map}{MAP}{maximum a posteriori}

\newtheorem{theorem}{Theorem}

\def\BibTeX{{\rm B\kern-.05em{\sc i\kern-.025em b}\kern-.08em
    T\kern-.1667em\lower.7ex\hbox{E}\kern-.125emX}}

\fancypagestyle{cfooter}{ %
\fancyhf{} 
	\cfoot{\scriptsize{This work has been submitted to the IEEE for possible publication. Copyright may be transferred without notice, after which this version may no longer be accessible.}
	}

}

\begin{document}

\bstctlcite{IEEEexample:BSTcontrol}

\title{Leveraging Variational Autoencoders for Parameterized MMSE Estimation
\thanks{This work is funded by the Bavarian Ministry of Economic Affairs, Regional Development, and Energy within the project 6G Future Lab Bavaria. The authors acknowledge the financial support by the Federal Ministry of Education and Research of Germany, project ID: 16KISK002.}
}

\author{\IEEEauthorblockN{Michael Baur,~\IEEEmembership{Graduate Student Member,~IEEE}, Benedikt Fesl,~\IEEEmembership{Graduate Student Member,~IEEE}, \\and Wolfgang Utschick,~\IEEEmembership{Fellow,~IEEE}}
}

\maketitle
\begin{abstract}
In this manuscript, we propose to use a variational autoencoder-based framework for parameterizing a conditional linear minimum mean squared error estimator. The variational autoencoder models the underlying unknown data distribution as conditionally Gaussian, yielding the conditional first and second moments of the estimand, given a noisy observation. The derived estimator is shown to approximate the minimum mean squared error estimator by utilizing the variational autoencoder as a generative prior for the estimation problem. We propose three estimator variants that differ in their access to ground-truth data during the training and estimation phases. The proposed estimator variant trained solely on noisy observations is particularly noteworthy as it does not require access to ground-truth data during training or estimation. We conduct a rigorous analysis by bounding the difference between the proposed and the minimum mean squared error estimator, connecting the training objective and the resulting estimation performance. Furthermore, the resulting bound reveals that the proposed estimator entails a bias-variance tradeoff, which is well-known in the estimation literature. As an example application, we portray channel estimation, allowing for a structured covariance matrix parameterization and low-complexity implementation. Nevertheless, the proposed framework is not limited to channel estimation but can be applied to a broad class of estimation problems. Extensive numerical simulations first validate the theoretical analysis of the proposed variational autoencoder-based estimators and then demonstrate excellent estimation performance compared to related classical and machine learning-based state-of-the-art estimators.
\end{abstract}

\section{Introduction}
\label{sec:intro}

\Acp{gm} are a class of \ac{ml} techniques designed to learn data distributions based on samples~\cite{Ruthotto2021}.
Instances of \acp{gm} are the \ac{gmm}~\cite[Ch.~9]{Bishop2006}, \ac{vae}~\cite{Rezende2014, Kingma2014}, \ac{gan}~\cite{Goodfellow2020}, and, more recently, the score-based model~\cite{Song2020}.
A trained \ac{gm} can generate unseen samples from the data distribution and often enables likelihood evaluation by providing a so-called generative prior. 
The generative prior is an approximation of the true data distribution and can be leveraged to solve sophisticated tasks such as inverse problems~\cite{Ongie2020}, dealing with recovering not directly observable parameters based on their noisy observations. 
Exemplarily, \cite{Bora2017} uses a \ac{gan} for image reconstruction by adopting a \ac{cs} framework and using the \ac{gan} as a generator. 
The approach is further extended in~\cite{Jalal2021} to MRI images. 
Other image processing-related examples solving inverse problems with \acp{gm} involve phase retrieval~\cite{Hand2018} and blind image deconvolution~\cite{Asim2020}.
In the context of wireless communications, generative priors find application in \ac{ce}~\cite{Koller2022, Balevi2021, Balevi2021a, Doshi2022, Arvinte2023}, where the channel is estimated based on a noisy pilot observation, representing another instance of an inverse problem.

For the solution of an estimation task, a frequentist framework assumes the data to be deterministic and commonly constrains the estimator class to be unbiased in search for a minimum variance unbiased estimator~\cite{Diskin2023}. 
In opposition, if a Bayesian approach is considered, it is well-known in estimation theory that the \ac{cme} delivers \ac{mmse} estimates~\cite[Ch.~10]{Kay1993}. 
Therefore, a (parameterized) Bayesian estimator's goal should be approximating the \ac{cme}. 
Moreover, a fundamental aspect of the Bayesian framework is modeling the data as a \ac{rv}, enabling the incorporation of a prior distribution into the estimation process. 
The result is an excellent estimation performance if the prior distribution accurately models the data, e.g., in the form of a generative prior. 
Therefore, \acp{gm} and Bayesian inference can be ideally combined to perform estimation tasks due to the distribution modeling abilities of the \acp{gm}.

A well-known \ac{gm} that can be used for directly approximating the \ac{cme} is the \ac{gmm}~\cite{Koller2022}. 
However, connections between the \ac{cme} and other \acp{gm} are yet to be discovered in the literature. 
Exemplarily, the \ac{gan}-based estimator from~\cite{Balevi2021, Balevi2021a, Doshi2022} is used in a \ac{cs}-fashioned way and the score-based approach from~\cite{Arvinte2023} requires an iterative posterior sampling process, causing a massive computational complexity. 
A closely related \ac{gm} to the \ac{gmm} is the \ac{vae}. 
Both \acp{gm} maximize a lower bound to the data log-likelihood and introduce an artificial latent space. 
Nevertheless, the \ac{gmm} utilizes a discrete latent space, which limits its expressiveness. 
On the contrary, a \ac{vae} uses a continuous latent space, resulting in a better representation ability and a more flexible architectural design. 
The \ac{vae} intrinsically makes no assumption about the data distribution and was shown to work well in domains where it is traditionally challenging to derive statistical data models, e.g., in image processing~\cite{Zhao2017a}.

In this work, we propose a \ac{vae}-parameterized estimator, combining a \ac{gm} and classical estimation theory, with the following \textit{contributions}: 
\begin{itemize}
    \item We model the analytically intractable data distribution as \ac{cg} with the help of the \ac{vae}, yielding conditional first and second moments to parameterize conditional \ac{lmmse} estimators given the latent representation and noisy observations. The conditional \ac{lmmse} estimators are \ac{mse}-optimal and analytically tractable in closed-form due to the \ac{cg} likelihood model.
    \item Since the \ac{vae} inherently makes no assumptions about the data distribution, the proposed estimation framework works independently of the adopted data distribution.
    \item We introduce a low-complexity estimator version based on a \ac{map} estimate requiring only one \ac{dnn} forward pass (\mapvae estimator). In contrast to many existing \ac{gm}-based estimation frameworks for inverse problems, e.g.~\cite{Balevi2021, Balevi2021a, Doshi2022, Arvinte2023}, this procedure allows for a computationally efficient approximation of the \ac{cme}, which is high-performing and robust, as a consequence of the \ac{vae} serving as a generative prior. Compared to classical algorithms for inverse problems such as \ac{ls} or \ac{amp}, the proposed \mapvae estimator achieves significant performance gains. 
    \item Three estimator variants are proposed, differing in the availability of ground-truth data during their training and estimation phases. The \textit{VAE-real} variant is particularly noteworthy as it requires no access to samples of ground-truth data in either the training or estimation phases. 
    \item We rigorously derive a bound on the performance gap between the \mapvae estimator and the \ac{cme}, allowing for an interpretable estimation procedure. The bound connects the training objective of the VAE with the resulting estimation performance and reveals that the proposed estimator entails a bias-variance tradeoff that is well-known in the estimation literature.
    \item As an application example, we consider \ac{ce}, offering a low-complexity implementation due to the structural properties of the \ac{ce} problem. Our extensive numerical simulations first validate the theoretical analysis and then underline the superiority of the proposed \ac{vae}-based estimator variants compared to the baseline methods under various system configurations. 
\end{itemize}

Moreover, we provide the following extensions in this manuscript compared to the preliminary results in~\cite{Baur2022}. 
The analyses in Sections~\ref{subsec:mmse-est-vae} and~\ref{subsec:convergence} enhance the theoretical foundation of the \ac{vae}-based estimator's \ac{mse}-optimality. 
We provide a more general treatment by providing a scheme for linear inverse problems of which \ac{mimo}-\ac{ce} is a special instance. 
We make the training of the \ac{vae} \ac{snr}-independent, meaning that we use a single trained \ac{vae} for every \ac{snr} value, in opposition to~\cite{Baur2022}, where an individual \ac{vae} is trained for every \ac{snr} value. 
Finally, the numerical simulations in this manuscript are more comprehensive.

The structure of this manuscript is as follows.
Section~\ref{sec:system} discusses the signal model and the general problem formulation and provides background information about the \ac{vae}. 
In Section~\ref{sec:vae-ce}, we introduce the \ac{vae}-based estimator and its three variants, followed by the derivation and interpretation of the error bound between the proposed estimator and the \ac{cme}. 
We discuss \ac{ce} as an application example in Section~\ref{sec:practical}. 
Numerical simulation results are presented in Section~\ref{sec:results}, and we conclude this manuscript in Section~\ref{sec:conclusion}.

\textit{Notation}: We denote vectors and matrices as lower-case and upper-case bold-faced symbols, respectively. 
Element-wise multiplication is denoted as $\odot$, the all-zeros vector as $\bm{0}$, and the all-ones vector as $\bm{1}$.
The vectorization operation $\vect(\mg)\in\CC^{g_1 g_2}$ stacks the columns of $\mg\in\CC^{g_1\times g_2}$ into a vector.
The Kronecker product of two matrices $\mb\in\CC^{b_1 \times b_2}$ and $\md\in\CC^{d_1 \times d_2}$ is $(\mb \otimes \md) \in \CC^{b_1 d_1 \times b_2 d_2}$.

\section{Preliminaries}
\label{sec:system}


\subsection{Signal Model and Problem Formulation}
\label{subsec:signal-problem}

We consider the generic linear inverse problem
\begin{equation}
    \vy = \ma \vh + \vn
    \label{eq:system-inv_prob}
\end{equation}
with the observation matrix $\ma\in\CC^{M \times N}$ and additive noise $\vn \sim \mathcal{N}_\CC (\bm 0, \msig)$.
It is assumed that the matrix $\ma$ and the noise covariance $\msig$ are given. 
The task is to recover $\vh$ based on $\vy$. 
The design of $\ma$ is characteristic of the problem to be solved, e.g., in \ac{ce}, $\ma$ represents the pilot allocation~\cite{Koller2022, Balevi2021, Balevi2021a, Doshi2022}. 
For further examples, we refer to~\cite{Rani2018}.

For the solution of~\eqref{eq:system-inv_prob}, we aim to estimate $\vh$ based on the noisy observation $\vy$.
In the Bayesian framework, $\vh$ is a \ac{rv} with an unknown prior $p(\vh)$. 
The goal is to minimize the \ac{mse}
\begin{equation}
    \label{eq:bmse}
    \E\left[ \| \vh - \hat\vh \|^2 \right] = \int \left[ \int \| \vh - \hat\vh \|^2 p(\vh\cnd\vy) \, \diff\vh \right] p(\vy) \, \diff\vy
\end{equation}
with the estimate $\hat\vh\in\CC^N$.
For minimizing the MSE, minimizing the inner integral is sufficient due to $p(\vy)\geq 0$. 
The minimizer is the well-known \ac{cme} 
\begin{equation}
    \E[\vh\cnd\vy] = \argmin_{\hat\vh}\E\left[ \| \vh - \hat\vh \|^2 \right]
    \label{eq:cme-mse}
\end{equation}
resulting in \ac{mse}-optimal estimates, cf.~\cite[Ch.~10]{Kay1993} for details.
More generally, the \ac{cme} is the optimal predictor for all Bregman loss functions, of which the \ac{mse} is a special case~\cite{1459065}. 
Application of Bayes' rule to $p(\vh\cnd\vy)$ yields
\begin{equation}
    \E[\vh\cnd\vy] =  \int \vh \frac{p_\vn(\vy - \ma\vh)\, p(\vh)}{p(\vy)} \mathrm{d} \vh.
    \label{eq:cme-bayes}
\end{equation}
Note that $p_\vn$ represents the noise \ac{pdf}.
By inspection of~\eqref{eq:cme-bayes}, it becomes clear why the \ac{cme} is difficult to compute. 
First, it requires access to the unknown and difficult-to-determine prior $p(\vh)$, necessitating an estimate of $p(\vh)$. 
Second, an approximation of the integral in~\eqref{eq:cme-bayes} is required since, in general, there exists no closed-form solution. 
Another approach may involve directly approximating $p(\vh\cnd\vy)$, e.g., with Monte-Carlo sampling methods. 
Nevertheless, this would, in general, still necessitate calculating an intractable integral over $p(\vh\cnd\vy$) to yield $\E[\vh\cnd\vy]$. 
Consequently, such procedures' applicability would be limited, especially in time-sensitive applications.


\subsection{VAE Fundamentals}
\label{subsec:vae}

In a parametric approach, the parameterized likelihood model $p_\vtheta(\vh)$ approximates the unknown prior $p(\vh)$. 
One of the simplest parametric models is assuming a Gaussian prior, parameterized with the sample mean and covariance. 
The resulting parameterized \ac{cme} approximation is the well-known \ac{lmmse} estimator~\cite[Ch. 10]{Kay1993}. 
However, assuming a Gaussian prior is restrictive, causing the estimator to perform weakly if the true prior strongly deviates from a Gaussian distribution, which is the case in real-world systems. 
A way to significantly improve the expressiveness of the likelihood model while preserving the favorable properties of a Gaussian distribution is to let it hold only conditionally so the data is modeled as \ac{cg}. 
The \ac{cg} likelihood model has the form 
\begin{equation}
    \vh \mid \vz \sim p_\vtheta(\vh\cnd\vz) = \mathcal{N}_{\CC}(\vmu_\vtheta(\vz), \mc_\vtheta(\vz))
    \label{eq:cg-vae}
\end{equation}
with the so-called latent vector $\vz\in\RR^\NL$ such that
\begin{equation}
    p_\vtheta(\vh) = \int p_\vtheta(\vh\cnd\vz) p(\vz) \diff \vz
    \label{eq:p_h}
\end{equation}
with a fixed $p(\vz)$.
Besides its great properties in terms of expressiveness, the \ac{cg} model in~\eqref{eq:cg-vae} will be a key aspect for deriving the \ac{vae}-parameterized estimator in Section~\ref{subsec:mmse-est-vae}.
Since $p_\vtheta(\vh\cnd\vz)$ is defined according to~\eqref{eq:cg-vae}, $\vtheta$ also implicitly parameterizes the intractable posterior 
\begin{equation}
    p_\vtheta(\vz\cnd\vh) = \frac{p_\vtheta(\vh\cnd\vz) p(\vz)}{\int p_\vtheta(\vh\cnd\vz) p(\vz) \mathrm{d} \vz}.
    \label{eq:posterior}
\end{equation}

\begin{figure}[t]
    \centering
    \begin{tikzpicture}[->, >=stealth]

    \def\nodeSep{50pt}
    \def\minSize{0.8cm}
    \def\innerSep{0.5em}
    
    \node[circle, draw, minimum size=\minSize, inner sep=\innerSep] (z) {$\bm{z}$};
    \node[circle, draw, minimum size=\minSize, inner sep=\innerSep] (h) [right=\nodeSep of z] {$\bm{h}$};
    \node[circle, draw, minimum size=\minSize, inner sep=\innerSep] (y) [right=\nodeSep of h] {$\bm{y}$};	
    
    \path (z) edge [solid] node [below]{$p_\vtheta(\vh\cnd\vz)$} (h);
    \path (h) edge [solid] node [below]{$p(\vy\cnd\vh)$} (y);
    \path (y) edge [bend right, dashed] node [above]{$q_\vphi(\vz\cnd\vy)$} (z);
    
\end{tikzpicture}
    \caption{Bayesian network illustrating the VAE decoder distribution $p_\vtheta(\vh\cnd\vz)$, encoder distribution $q_\vphi(\vz\cnd\vy)$, and the known $p(\vy\cnd\vh)=\mathcal{N}_\CC(\ma\vh, \mathbf{\Sigma})$.}
    \label{fig:prob_graph}
\end{figure}

A blueprint to obtain $\vtheta$ is given by the Bayesian network in Fig.~\ref{fig:prob_graph} that parameterizes the joint \ac{pdf}
\begin{equation}
    p_\vtheta(\vy,\vh,\vz) = p(\vy\cnd\vh)\, p_\vtheta(\vh\cnd\vz)\, p(\vz) .
    \label{eq:joint_vae}
\end{equation}
Since the system model in~\eqref{eq:system-inv_prob} sets $p(\vy\cnd\vh) = \mathcal{N}_\CC(\ma\vh, \mathbf{\Sigma})$, only $p_\vtheta(\vh\cnd\vz)$ must be learned in~\eqref{eq:joint_vae} if $p(\vz)$ is fixed.
By inspection of Fig.~\ref{fig:prob_graph}, the following becomes apparent: all involved \acp{rv} are stochastically dependent, whereas $\vy$ and $\vz$ are conditionally independent given $\vh$ due to the local Markov property in the Bayesian network.
Moreover, the learnable $q_\vphi(\vz\cnd\vy)$ in Fig.~\ref{fig:prob_graph} symbolizes the connection from $\vy$ to $\vz$, which we will use to infer a $\vz$ based on $\vy$ in Section~\ref{subsec:mmse-est-vae} because the true posterior $p_\vtheta(\vz\cnd\vy)$ is intractable, cf.~\eqref{eq:posterior}.

\begin{figure}[t]
      \centering
      \begin{tikzpicture}[>=stealth, scale=0.9]
	\def\NNsize{2cm}
	\def\NNwidth{2cm}
	\def\NNheight{2.2cm}
	\def\NNangle{70}
	\node (input) at (0, 0) { $\bm{y}$};
	
	\node[trapezium, draw, align=center, trapezium stretches = true, minimum height=\NNwidth, minimum width=\NNheight, align=center, trapezium angle=\NNangle, rotate=-90] (NN1) at($ (input) + (2,0) $) {\rotatebox{90}{Encoder} \small{ \rotatebox{90}{\hspace{-1mm} $q_{\vphi}(\bm{z}\cnd\bm{y})$}}};
	
	\node[circle, draw, inner sep = 0.01em] (add) at ($ (NN1.north) + (1.3,0.4) $) {$ + $};
	\node[circle, draw, inner sep = 0.01em] (multi) at ($ (NN1.north) + (1.3,-0.4) $) {$ \odot $};
	\node (eps) at ( $(multi) + (0.3, -1)$ ) {\quad\quad\quad $\bm{\varepsilon} \sim \mathcal{N}(\bm{0},\mathbf{I}) $};
	
	\node[trapezium, draw, align=center, trapezium stretches = true, minimum height=\NNwidth, minimum width=\NNheight, align=center, trapezium angle=\NNangle, rotate=90](NN2) at ($ (NN1) + (4.6,0) $) {\small{\rotatebox{-90}{\hspace{0.25mm}$p_{\vtheta}(\bm{h}\cnd\bm{z})$}}\hspace{1mm} \normalsize{\rotatebox{-90}{Decoder}}};
	
	
	\node(output_mu) at ($ (NN2.south) + (1.3,0.4) $) { $\vmu_\vtheta(\vz)$};
	\node(output_cov) at ($ (NN2.south) + (1.3,-0.4) $) { $\mc_\vtheta(\vz)$};
	
	\draw[->] (input.east) -- (NN1.south) {};
	
	\draw[->] ($ (NN1.north) + (0,0.4) $) --node[midway, above=-.0em]{\hspace{-0.2mm} $\vmu_\vphi(\vy)$}    (add.west);
	\draw[->] ($ (NN1.north) + (0,-0.4) $) --node[midway, above=-1.9em]{ $\vsig_\vphi(\vy)$} (multi.west);
	\draw[->] (multi.north) -- (add.south);
	\draw[->] ($ (eps.north) - (0.3, 0.1) $)   -- (multi.south);
	\draw[->] (add.east) --node[midway, above=.1em]{ $\bm{z}$} ($(NN2.north) + (0,0.4)$);
	
	\draw[->] ($ (NN2.south) + (0,0.4) $) -- (output_mu.west);
	\draw[->] ($ (NN2.south) + (0,-0.4) $) -- (output_cov.west);
\end{tikzpicture}
    \vspace{-7mm}
    \caption{Structure of a VAE with \ac{cg} distributions for $q_{\vphi}(\vz\cnd\vy)$ and $p_{\vtheta}(\vh\cnd\vz)$. The encoder and decoder each represent a \ac{nn}.}
    \label{fig:vae}
    \vspace{-2mm}
\end{figure}

The \ac{vae}~\cite{Kingma2014} practically realizes the considered Bayesian network, for which an illustration is visible in Fig.~\ref{fig:vae}.
For the \ac{vae} training, $p_\vtheta(\vh)$ is typically decomposed as~\cite{Kingma2019}
\begin{equation}
    \log p_\vtheta(\vh) = \mathcal{L}_{\vtheta,\vphi}(\vh) + \KL(q_\vphi(\vz\cnd\vy)\,\|\,p_\vtheta(\vz\,|\,\vh))
    \label{eq:log-like}
\end{equation}
with the \ac{elbo}
\begin{equation}
    \mathcal{L}_{\vtheta,\vphi}(\vh) = \E_{q_{\vphi}} \left[\log p_{\vtheta}(\vh\cnd\vz)\right] - \KL(q_{\vphi}(\vz\cnd\vy)\,\|\,p(\vz))
    \label{eq:vae}
\end{equation}
and the non-negative \ac{kl} divergence 
\begin{equation}
    \KL(q_\vphi(\vz\cnd\vy)\,\|\,p_\vtheta(\vz\cnd\vh)) = \E_{q_\vphi}\left[ \log \left( \frac{q_\vphi(\vz\cnd\vy)}{p_\vtheta(\vz\cnd\vh)} \right) \right].
    \label{eq:elbo-gap}
\end{equation}
Note that $\E_{q_\vphi(\vz|\vy)}[\cdot] = \E_{q_\vphi}[\cdot]$.
The variational distribution $q_\vphi(\vz\cnd\vy)$ is introduced aiming to approximate the intractable $p_\vtheta(\vz\cnd\vh)$ as can be seen in~\eqref{eq:elbo-gap}.
Contrary to the \ac{vae} from~\cite{Kingma2014}, the variational distribution here is conditioned on $\vy$ instead of $\vh$ since the latter will be inaccessible during the estimation phase after the training.
Consequently, a maximization of the \ac{elbo} is independent of~\eqref{eq:posterior}, maximizes $\log p_\vtheta(\vh)$, as well as minimizes~\eqref{eq:elbo-gap}.
In summary, a sufficiently trained \ac{vae} yields $\vtheta$ for the \ac{cg} model in~\eqref{eq:cg-vae}, as well as an approximation of the intractable posterior in~\eqref{eq:posterior} via $q_\vphi(\vz\cnd\vy)$.

The remaining distributions in~\eqref{eq:vae} are defined as:
\begin{align}
    q_{\vphi}(\vz\cnd\vy) &= \mathcal{N}(\vmu_\vphi(\vy),\diag(\vsig^2_\vphi(\vy))),    \label{eq:q_phi}\\
    p(\vz) &= \mathcal{N}(\bm{0},\mathbf{I}).    \label{eq:prior}
\end{align}
Moreover, the \ac{vae} implements $p_\vtheta(\vh\cnd\vz)$ and $q_\vphi(\vz\cnd\vy)$ as \acp{dnn}.
With these considerations, let us revisit Fig.~\ref{fig:vae}.
The encoder takes an observation $\vy$ and maps it to $\vmu_\vphi(\vy)$ and $\vsig_\vphi(\vy)$ to obtain a reparameterized sample $\vz= \vmu_\vphi(\vy)+\vsig_\vphi(\vy)\,\odot\,\veps$. 
The sample $\vz$ is fed into the decoder to obtain $\vmu_\vtheta(\vz)$ and $\mc_\vtheta(\vz)$ representing the first and second moments of $p_{\vtheta}(\vh\cnd\vz)$.

Due to the \ac{cg} distributions, the terms in the \ac{elbo} can be calculated analytically, which is beneficial for the training of the \ac{vae}.
The expectation term in~\eqref{eq:vae} can be approximated with a single sample $\tilde\vz\sim q_{\vphi}(\vz\cnd\vy)$, i.e., $(-\E_{q_{\vphi}} \left[\log p_{\vtheta}(\vh\cnd\vz)\right])$ is replaced by the estimate
\begin{equation}
     \log\det(\pi\,\mc_\vtheta(\tilde\vz)) + (\vh - \vmu_\vtheta(\tilde\vz))\herm \mc^{-1}_\vtheta(\tilde\vz) (\vh - \vmu_\vtheta(\tilde\vz)).
    \label{eq:vae-dec-like}
\end{equation}
The KL divergence $\KL(q_{\vphi}(\vz\cnd\vy)\,\|\,p(\vz))$ in~\eqref{eq:vae} results in
\begin{equation}
    \frac{1}{2} \left( \bm{1}\tran \left( -\log\vsig^2_\vphi(\vy) + \vmu^2_\vphi(\vy) + \vsig^2_\vphi(\vy) \right) - \NL \right).
    \label{eq:vae-kl}
\end{equation} 
By utilizing an expressive decoder \ac{dnn} and $\mc_\vtheta(\vz)$ parameterization, we assume that a properly trained \ac{vae} where~\eqref{eq:cg-vae} holds delivers a $p_\vtheta(\vh)$ that well approximates $p(\vh)$. 
We will explicitly discuss conditional covariance matrix parameterizations for $\mc_\vtheta(\vz)$ in Sections~\ref{subsec:covariance} and~\ref{sec:practical}.

\section{VAE-Parameterized Estimator}
\label{sec:vae-ce}


\subsection{MMSE Estimation with the VAE}
\label{subsec:mmse-est-vae}

After its successful training, the \ac{vae} yields $p_\vtheta(\vh\cnd\vz)$ as \ac{cg} according to~\eqref{eq:cg-vae}.
Recall the corresponding Bayesian network in Fig.~\ref{fig:prob_graph} visualizing the dependencies of the involved \acp{rv}, which will be helpful for the following estimator derivation. 
Starting from~\eqref{eq:cme-bayes}, the law of total expectation enables reformulating the \ac{cme} as~\cite[Sec. 4.3]{Bertsekas2014}:
\begin{equation}
    \E[\vh\cnd\vy] =  \E_{p_{\vtheta}(\vz\cnd\vy)}[\E_\vtheta[\vh\cnd\vz,\vy]\cnd\vy],
    \label{eq:total_exp}
\end{equation}
where the inner expectation is with respect to $p_\vtheta(\vh\cnd\vz,\vy)$. 
We neglect a possible approximation error between $p_\vtheta(\vh)$ and $p(\vh)$ in~\eqref{eq:total_exp} as it is irrelevant for the estimator derivation. 
Since $p(\vh)$ is anyway inaccessible, an analysis of such an error is only possible empirically in terms of an \ac{mse} investigation, which will be done in Section~\ref{sec:results}. 
Similar to $p_\vtheta(\vz\cnd\vh)$, $p_\vtheta(\vz\cnd\vy)$ and $p_\vtheta(\vh\cnd\vz,\vy)$ are also implicitly parameterized by $\vtheta$ due to~\eqref{eq:posterior}, the fixed prior $p(\vz)$ in~\eqref{eq:prior} and the model in~\eqref{eq:system-inv_prob}.
Indeed, $p_\vtheta(\vz\cnd\vy)$ is generally inaccessible for the same reason as $p_\vtheta(\vz\cnd\vh)$, cf.~\eqref{eq:posterior}. 
Since the encoder receives $\vy$ as input and $\vh$ is conditioned on $\vz$ according to~\eqref{eq:cg-vae}, the training objective in~\eqref{eq:vae} enforces $\vn$ and $\vz$ to be independent. 
Then, given~\eqref{eq:cg-vae}, we obtain a closed-form expression for the inner expectation in~\eqref{eq:total_exp} due to the \ac{cg} property causing $\vy$ and $\vh$ to be jointly Gaussian given~$\vz$. 
Therefore, $\E_\vtheta[\vh\cnd\vz,\vy]$ results in~\cite{Yang2015}:
\begin{equation}
    \vmu_\vtheta(\vz) + \mc_\vtheta(\vz) \ma\herm (\ma\mc_\vtheta(\vz)\ma\herm + \msig)^{-1} (\vy - \ma\vmu_\vtheta(\vz)),
    \label{eq:lmmse-vae}
\end{equation}
where the matrix $\ma$ and vector $\vy$ belong to~\eqref{eq:system-inv_prob}, and $\vmu_\vtheta(\vz)$, $\mc_\vtheta(\vz)$, and $\vz$ to~\eqref{eq:cg-vae}.
See Appendix~\ref{app:derivation-lmmse} for a step-by-step derivation of~\eqref{eq:lmmse-vae}.

It remains to solve the intractable outer expectation in~\eqref{eq:total_exp}. 
To this end, the approximation of $p_\vtheta(\vz\cnd\vh)$ via $q_\vphi(\vz\cnd\vy)$ in~\eqref{eq:elbo-gap} comes into play.
Although~\eqref{eq:elbo-gap} shows the approximation of $p_\vtheta(\vz\cnd\vh)$ instead of $p_\vtheta(\vz\cnd\vy)$, the parameter combination that maximizes $p_\vtheta(\vh\cnd\vz)$ in~\eqref{eq:vae} also maximizes $p_\vtheta(\vy\cnd\vz)$ since the noise distribution is considered to be known and not subject to optimization, thus permitting the substitution.
Consequently, by replacing $p_\vtheta(\vz\cnd\vy)$ with $q_\phi(\vz\cnd\vy)$ in~\eqref{eq:total_exp}, 
\begin{equation}
    \E[\vh\cnd\vy] \approx \E_{q_\vphi} \left[ t_\vtheta(\vz,\vy) \right], \quad t_\vtheta(\vz,\vy) = \E_\vtheta[\vh\cnd\vz,\vy].
    \label{eq:vae-est-q}
\end{equation}
As we can easily obtain samples of $q_\phi(\vz\cnd\vy)$ with the help of the encoder, we can approximate $\E[\vh\cnd\vy]$ using samples of the form $ \vz^{(k)} = \vmu_\vphi(\vy) + \bm{\varepsilon}^{(k)} \odot \vsig_\vphi(\vy) $ where every $ \bm{\varepsilon}^{(k)} $ is a sample from $ \mathcal{N}(\bm{0},\mathbf{I}), k=1,\ldots,K$.
Based on the samples $\vz^{(k)}$ we can approximate the \ac{mmse} estimator as a consequence of the law of large numbers~\cite{Loeve1977}:
\begin{equation}
    \hat\vh_{\text{VAE}}^{(K)}(\vy) = \frac{1}{K} \sum_{k=1}^K t_\vtheta(\vz^{(k)},\vy), \quad \vz^{(k)} \sim q_\phi(\vz\cnd\vy),
    \label{eq:vae-estimator}
\end{equation}
where $t_\vtheta(\vz^{(k)},\vy)$ is evaluated with~\eqref{eq:lmmse-vae}.

The estimator $\hat\vh_{\text{VAE}}^{(K)}(\vy)$ generally becomes better for a large number of samples $K$.
However, a large $K$ is unwanted in a real-time system. 
It is desirable to reduce the complexity of the estimator as much as possible, which means that $K$ should be low. 
To this end, we first obtain a \ac{map} estimate for $\vz$, which is $\vmu_\vphi(\vy)$ at the encoder output due to the Gaussianity of $q_\vphi(\vz\cnd\vy)$, cf.~\eqref{eq:q_phi}. 
The \ac{map} estimate is subsequently passed in a single step through the decoder to evaluate $t_\vtheta(\vz,\vy)$.
Consequently, we define the \mapvae estimator 
\begin{equation}
    \hat\vh_{\text{VAE}}(\vy) = \hat\vh_{\text{VAE}}^{(1)}(\vy) = t_\vtheta( \vz^{(1)}=\vmu_\vphi(\vy), \vy)
    \label{eq:vae-estimator-mu}
\end{equation}
based on the \ac{map} estimate $\vz^{(1)}=\vmu_\phi(\vy)$ from $q_\vphi(\vz\cnd\vy)$. 
In Section~\ref{subsec:convergence}, we rigorously analyze the performance gap between the \mapvae estimator and the \ac{cme}.
Furthermore, in Section~\ref{subsec:res_arcstudy}, we compare $\hat\vh_{\text{VAE}}^{(K)}(\vy)$ and $\hat\vh_{\text{VAE}}(\vy)$ for different $K$, demonstrating that their estimation quality is nearly identical.
Unless otherwise stated, we use the \mapvae estimator in~\eqref{eq:vae-estimator-mu} for the numerical simulations.


\subsection{Covariance Matrix Parameterization}
\label{subsec:covariance}

According to~\eqref{eq:cg-vae}, the \ac{vae} aims to learn a full covariance matrix for $\vh\cnd\vz$. 
However, learning such a full matrix requires learning a large number of parameters, resulting in huge \acp{dnn}. 
It is also known that covariances usually exhibit problem-specific structures, which can be exploited to drastically reduce the number of parameters to learn.

In this work, we consider equidistantly sampled \ac{wss} random processes covering a broad class of applications in signal processing:
\begin{itemize}
    \item sensor array processing with equidistantly spaced sensors, e.g., for beamforming \cite{Fuhrmann1991} or speech processing~\cite{Ephraim1989}
    \item \ac{ce} in the spatial and time-frequency domain~\cite{Koller2022, Neumann2018}
    \item times series analysis in financial markets~\cite{Heckens2020}
\end{itemize}
As a result of the \ac{wss} assumption, the covariance matrix is Toeplitz structured. 
The parameterization of a Toeplitz matrix is possible with an oversampled \ac{dft} matrix as demonstrated in~\cite{Fesl2022, Baur2023, Baur2023meas}. 
However, if the covariance matrix dimensions are large, a circulant matrix can asymptotically approximate the Toeplitz covariance matrix~\cite{Gr06}. 
By reasonably assuming that the VAE finds latent conditions that preserve the structural properties of the second moments~\cite{Bock2024}, we can choose
\begin{equation}
    \mc_\vtheta(\vz)=\mf^\mathrm{\,H}_N\diag(\vc_\vtheta(\vz))\mf_N, \qquad \vc_\vtheta(\vz)\in\RR_+^N,
    \label{eq:circulant}
\end{equation}
parameterizing a circulant matrix, where $\mf_N\in\CC^{N \times N}$ is a \ac{dft} matrix. 
By choosing a covariance matrix parameterization in accordance with the structure of the actual covariance, assuming $p_\vtheta(\vh\cnd\vz)$ well-approximates the true distribution is reasonably motivated.

Circulant matrices have the advantage that they allow for a low-complexity and memory-efficient implementation and have already been used in previous work, cf.~\cite{Neumann2018}. 
This can be seen in~\eqref{eq:circulant} since a positive and real-valued vector $\vc_\vtheta(\vz)$ suffices to parameterize a full covariance matrix. 
Due to the \ac{dft} matrix, \eqref{eq:circulant} is furthermore straightforwardly invertible in $\OO(N\log N)$ time (by using FFTs), motivating its usage in the proposed \ac{vae}-based estimation framework.


\subsection{Variants of VAE-based Estimators}
\label{subsec:vae-est-variants}

We present three possible estimator variants that leverage the \ac{vae}. 
All three estimators have in common that the \acp{vae} can be trained offline before application. 
The estimators differ in their ground-truth data knowledge during the training and evaluation phase. 
A comprehensive overview of all variants with their losses will be shown in Section~\ref{sec:practical} in Table~\ref{tab:vae-variants}.

\textit{1) VAE-genie}: 
To determine the full potential of our method, we assume $\vn=\bm{0}$ in~\eqref{eq:system-inv_prob} for the encoder input while \eqref{eq:lmmse-vae} is still evaluated with a non-zero noise realization. 
VAE-genie is supposed to exhibit the best estimation results among all variants because the $\vmu_\vtheta(\vz)$ and $\mc_\vtheta(\vz)$ at the decoder are inferred with the ground-truth data at the encoder and its latent representation. 
Although VAE-genie even has the potential to outperform the \ac{cme}, as the ground-truth data acts as side information, this estimator is not applicable in practice, since it requires ground-truth knowledge during the evaluation phase. 
Instead, it can be a suitable benchmark result in a scenario where the optimal estimator is unknown and inaccessible. 
VAE-genie requires ground-truth data knowledge during the training \textit{and} evaluation phase.

\textit{2) VAE-noisy}: 
This estimator version directly relates to Fig.~\ref{fig:vae}.
The encoder receives the noisy observation $\vy$ as input with $\vn\neq\bm{0}$. 
VAE-noisy only requires ground-truth data access during the training phase to compute~\eqref{eq:vae-dec-like} for its loss. 
During the evaluation phase, the mean $\vmu_\vphi(\vy)$ is obtained based on the noisy observation $\vy$ to compute~\eqref{eq:lmmse-vae}, which is the reason for the name of this estimator. 
We expect that VAE-noisy delivers worse estimation quality than VAE-genie as VAE-genie has ground-truth knowledge in the evaluation phase. 
VAE-noisy is, in contrast, applicable in practice.

\textit{3) VAE-real}: 
Similar to VAE-noisy, this estimator variant also receives $\vy$ as encoder input.
The change compared to VAE-noisy happens at the decoder in Fig.~\ref{fig:vae} where VAE-real learns first and second moments for $p_\vtheta(\vy\cnd\vz)$ instead of $p_\vtheta(\vh\cnd\vz)$. 
However, to efficiently compute $\E_\vtheta[\vh\cnd\vz,\vy]$ we require a \ac{cg} model for $\vh$ and not $\vy$.
As long as $\E[\vn]=\bm{0}$, which is the case in~\eqref{eq:system-inv_prob}, the mean of $\vy\cnd\vz$ is $\ma\vmu_\vtheta(\vz)$.
A simple workaround can determine the conditional covariance of $\vy\cnd\vz$.
While the VAE decoder continues to output $\mc_\vtheta(\vz)$, e.g., according to~\eqref{eq:circulant}, the matrix $\ma\mc_\vtheta(\vz)\ma\herm+\msig$ is used as covariance for $p_\vtheta(\vy\cnd\vz)$.
Consequently, in~\eqref{eq:vae-dec-like}, VAE-real replaces $\vmu_\vtheta(\vz)$ with $\ma\vmu_\vtheta(\vz)$ and $\mc_\vtheta(\vz)$ with $\ma\mc_\vtheta(\vz)\ma\herm+\msig$ during the training.
This way, the decoder forces to substitute only the desired part, the conditional covariance $\mc_\vtheta(\vz)$, which is used for the computation of~\eqref{eq:lmmse-vae}. 
It should be noted that no ground-truth data is needed by VAE-real, neither during training nor during evaluation. 
VAE-real is the most realistic estimator variant since noisy observations can be utilized to train the \ac{vae}. 
In contrast, access to ground-truth data during the training phase is usually related to a considerable additional effort and may sometimes be impractical.


\subsection{MSE-Optimality and Conditional Bias-Variance Tradeoff}
\label{subsec:convergence}

In this section, we provide a theoretical analysis of the introduced \mapvae estimator. Before establishing a bound on the difference between the \mapvae estimator and the \ac{cme}, let us denote the decoder \ac{nn} functions as 
\begin{align}
    f_{\vtheta,1} &: \RR^\NL \to \CC^N, \vz \mapsto \vmu_{\vtheta}(\vz),
    \\
    f_{\vtheta,2} &: \RR^\NL \to \mathcal{C}^N_+, \vz \mapsto \mc_{\vtheta}(\vz),
\end{align}
where $\mathcal{C}^N_+$ is the set of all $N\times N$ \ac{psd} matrices (we consider the case of a full covariance matrix as this trivially includes all parameterized covariances discussed in \Cref{subsec:covariance}).
In this section, we assume $\vy=\vh+\vn$ to analyze the theoretical properties independent of $\ma$.

\begin{theorem}\label{cor:cme_bound}
    Consider a decorrelated observation $\B y = \B h +~\B n$ with $\B n \sim \mathcal{N}_\CC(\B 0, \varsigma^2 \eye)$ and let \eqref{eq:cg-vae}
    and \eqref{eq:total_exp} hold.
    Further, assume the decoder neural network functions are Lipschitz continuous, i.e., for $i=\{1,2\}$ and $\va,\vb \in \RR^\NL$,
    \begin{align}
        \|f_{\vtheta,i}(\va) - f_{\vtheta,i}(\vb)\|_2 \leq L_i \|\va - \vb \|_2.
        \label{eq:lipschitz}
    \end{align}
    Then, the expected Euclidean distance between the \ac{cme} \eqref{eq:total_exp} and the \mapvae estimator \eqref{eq:vae-estimator-mu} is upper bounded as
    \begin{align}
    \begin{aligned}
        \E&\Bigl[\bigl\|\E[\vh \cnd \vy] - \hat{\vh}_{\textup{\text{VAE}}}(\vy) \bigr\|_2\Bigl] \leq (C_1L_1 + C_2L_2) 
        \\
        &\cdot\left(\sqrt{\tr(\mc_{p_{\vtheta}(\vz\cnd\vy)})} + \sqrt{\E\Bigl[\left\|\vmu_{p_{\vtheta}(\vz\cnd\vy)} - \vmu_{\vphi}(\B y)\right\|^2_2\Bigr]}\right)
        \label{eq:cme_bound_exp}
    \end{aligned}
    \end{align}
    with the \ac{snr}-dependent factors
    \begin{align}
        C_1 = \sqrt{\E \left[\frac{\varsigma^4}{(\xi_{\textup{\text{min}}}(\B y) + \varsigma^2)^2}\right]},~~~
        C_2 = \sqrt{\frac{N}{\varsigma^2}}.
        \label{eq:c1_c2_exp}
    \end{align}
     where $\xi_{\textup{\text{min}}}(\B y)$ is the smallest eigenvalue of $\mc_{\vtheta}(\vmu_{\vphi}(\B y))$.
\end{theorem}
\textit{Proof:} See Appendix \ref{app:cme_bound2}.

\subsubsection{MSE-Optimality}
Theorem \ref{cor:cme_bound} shows the expected distance of the \mapvae estimator to the \ac{cme} only depends on the first two moments of the posterior distribution $p_\btheta(\B z \cnd \B y)$, which is approximated by $q_\bphi(\B z \cnd \B y)$.
In particular, the bound is smaller if the first moments of $q_\bphi(\B z \cnd \B y)$ match those of $p_\btheta(\B z \cnd \B y)$, which can reasonably assumed to be the case after successfully training the \ac{vae}, being a mild assumption as no restrictions to higher moments apply.

Moreover, the smaller the variances of $p_{\btheta}(\B z \cnd \B y)$, the better the \ac{cme} approximation. Intuitively, this means that the less stochastic a mapping from the observation to the latent space is, the better the \mapvae estimator performs. 
Let us consider the following setup to motivate the encoder variances to become small after training. Assume the input data is compressible onto a lower-dimensional manifold, i.e., a lossless compression mapping exists from $\CC^N$ to $\RR^\NL$. In particular, this is known to be fulfilled for \textit{natural signals}, e.g., images or audio signals, wireless channels (especially in mmWave systems), or, in general, signals that exhibit a sparse representation through a dictionary. 
Then, a deterministic mapping exists into the latent space that can be learned by the \ac{vae}. In other words, there is no necessity for a stochastic mapping, and the variances in \eqref{eq:cme_bound_exp} can be chosen as zero without performance loss. 
This holds without restriction for the VAE-genie variant, where the encoder input is noiseless. 
For the VAE-noisy and VAE-real variants, although the latent encoding is trained to be stochastically independent of the noise, finding a deterministic mapping may be more intricate, especially in the low \ac{snr} regime, yielding a possibly higher encoder variance. 
We elaborate on this hypothesis in more detail for the example of channel estimation in \ac{mimo} systems in \Cref{sec:practical} and show through simulations in \Cref{subsec:res_arcstudy} that the \ac{vae}'s encoder variances are indeed converging towards zero during the training process.

Concluding the above discussion, the bound in Theorem~\ref{cor:cme_bound} establishes a connection between the training of the \ac{vae}, purely relying on likelihood maximization, and the resulting \ac{mse} performance. Moreover, the impact of the latent dimension on the estimation performance is better interpretable. 
Thus, by a successful training of a well-designed \ac{vae}, the resulting parameterized estimator converges to the \ac{cme}, thereby achieving a low \ac{mse}. We validate this proposition also through numerical results in \Cref{subsec:res_arcstudy}.

\subsubsection{Conditional Bias-Variance Tradeoff}
In addition to the above insights about the connection of the \ac{vae}'s training and the resulting estimation performance, the constants $C_1$ and $C_2$ in \eqref{eq:c1_c2_exp} have a reciprocal behavior over the \ac{snr} and, in particular, are vanishing in the high and low \ac{snr}, respectively. That is,
\begin{align}
    \lim_{\varsigma^2 \to 0} C_1= 0,~~~ \lim_{\varsigma^2 \to \infty} C_2 = 0.
\end{align}
Interestingly, this can be interpreted as a \textit{conditional bias-variance tradeoff} since $C_2 L_2$ in \eqref{eq:cme_bound_exp}, addressing the contribution of the conditional covariances, vanishes in low \ac{snr}; moreover, $C_1 L_1$, attributed to the conditional means, vanish in high \ac{snr}, cf.~\eqref{eq:cond-bias-variance}. 
Thus, the parameterized conditional covariance quality is less critical in the low SNR regime, as the parameterized LMMSE estimator relies more on the conditional first moment and vice versa in the high SNR regime. 
Consequently, the respective error terms have less impact on the bound to the CME. 
The entailment of such a \textit{conditional bias-variance tradeoff} is a highly desirable property of the proposed estimator as it serves as a regularization for the estimation performance and allows for great interpretability. 
Moreover, the analysis holds without restriction for all discussed estimator variants in \Cref{subsec:covariance} and all parameterized conditional covariance matrices since we made no assumptions about their structural properties.

\section{Example Application: Channel Estimation}
\label{sec:practical}

In this work, we consider \ac{mimo} \ac{ce} as an application example. 
In a \ac{mimo} communications system, the transmitter with $\Ntx$ antennas sends $\Np$ pilots to the receiver with $\Nrx$ antennas for estimating the channel matrix $\mh\in\CC^{\Nrx\times\Ntx}$. 
More precisely, the noisy observations 
\begin{equation}
    \my = \mh \mx + \mn \in \CC^{\Nrx\times\Np}
    \label{eq:system-mimo}
\end{equation}
are obtained at the receiver with the pilot matrix $\mx\in\CC^{\Ntx\times\Np}$ and noise matrix $\mn$.
After vectorizing~\eqref{eq:system-mimo}, the relation to~\eqref{eq:system-inv_prob} becomes apparent.
Consequently, $\vy=\vect(\my)$, $\vh=\vect(\mh)$, $\ma=(\mx\tran\otimes\,\mathbf{I})$, and $\vn=\vect(\mn)$.
Further, $M=\Nrx\Np$ and $N=\Nrx\Ntx$.
We investigate the uplink of a communications system where the \ac{mt} transmits to the \ac{bs} with $\Ntx<\Nrx$. 
However, the proposed framework can also be applied to the downlink since $\ma$ has a comparable structure.

We assume that the \ac{bs} and \ac{mt} are both equipped with a \ac{ula} with half-wavelength spacing. 
Note that a different array structure or antenna spacing can be straightforwardly reflected by the \ac{vae}'s parameterized covariance at the decoder output. 
Furthermore, we consider the fully determined case of~\eqref{eq:system-mimo}, i.e., $\Np=\Ntx$. 
We utilize \ac{dft} pilots, resulting in a unitary $\mx$, which results in a unitary $\ma$.
Moreover, we assume $\msig=\varsigma^2\mathbf{I}$ with given $\varsigma^2$.
Therefore, we perform a \ac{ls} estimate of \eqref{eq:system-inv_prob} to interpret it as a denoising task relating directly to the theoretical analysis in the previous section. 
The underdetermined case involving a wide $\ma$ is investigated in~\cite{Baur2023} and the \ac{ura} case at the \ac{bs} in~\cite{Baur2023meas} covering more advanced setups. 
The works~\cite{Baur2023, Baur2023meas} demonstrate a superior performance of the \ac{vae}-based estimators, highlighting the framework's versatile applicability under various system configurations.

Due to the common \ac{wss} assumptions in wireless communications~\cite[Sec. 2.6]{Yin2016}, the transmit- and receive-side covariance matrices at the BS and MT side, respectively, are Toeplitz structured, which are approximated by circulant matrices as explained in Section~\ref{subsec:covariance}. 
When additionally assuming uncorrelated scattering in the vicinity of the transmitter and receiver, we can decompose the \ac{ccm} into the Kronecker product of the transmit- and receive-side circulant-structured covariance matrices~\cite{Kermoal2002}:
\begin{equation}
    \mc_\vtheta(\vz) = \mq\herm \diag(\vc_\vtheta(\vz)) \mq, \quad \vc_\vtheta(\vz) \in \RR^N_+
    \label{eq:block-circ-vae}
\end{equation}
where $\mq = (\mf_\Ntx \otimes \mf_\Nrx)$.
In~\eqref{eq:block-circ-vae}, $\mc_\vtheta(\vz)$ is a block-circulant matrix, possessing the same favorable attributes regarding memory efficiency and low-complexity as an ordinary circulant matrix.
For a \ac{simo} system, which implies $\Ntx=1$, \eqref{eq:block-circ-vae} simplifies to~\eqref{eq:circulant}.


\subsection{Training Loss and Network Architecture}
\label{subsec:loss-architecture}

In principle, we train a \ac{vae} with the loss in~\eqref{eq:vae}, and, after the training, perform \ac{ce} as described in Section~\ref{subsec:mmse-est-vae}. 
Indeed, we can simplify~\eqref{eq:vae} as a result of the circulant parameterization. 
Let $\vh_Q=\mq\vh$, then the negative decoder likelihood in~\eqref{eq:vae-dec-like} can be expressed as
\begin{equation}
    N\log\pi + \bm{1}\tran \left( \bm{\lambda}_\vtheta(\tilde\vz) \odot |\vh_Q - \mq\vmu_\vtheta(\tilde\vz)|^2 - \log\bm{\lambda}_\vtheta(\tilde\vz) \right)
    \label{eq:dec-like-diag}
\end{equation}
with $\bm{\lambda}_\vtheta(\vz) = \vc^{-1}_\vtheta(\vz)$ and the element-wise absolute value~$|\cdot|$.
Eq.~\eqref{eq:vae-dec-like} reduces the numerical complexity during the training process because it avoids the inversion of a full covariance matrix compared to~\eqref{eq:dec-like-diag}. 
What is more, we utilize the \ac{ls} estimate of~\eqref{eq:system-inv_prob} (or $\vh$ for VAE-genie) multiplied with~$\mq$ as encoder input. 
Thus, the encoder input is effectively transformed into the angular or beamspace domain~\cite[Sec. 7.3]{Tse2005}, which is known to be sparse or highly compressible in massive \ac{mimo} systems, especially in mmWave systems. 
This validates the hypothesis of having a deterministic compression mapping that can be learned through the encoder in Section~\ref{subsec:convergence}. 
Therefore, a performance of the MAP-VAE estimator close to the \ac{cme} can be expected, which is also seen later in Section~\ref{sec:results}.

Combining every aspect from this section, the reformulated training objective that VAE-noisy is supposed to minimize is:
\begin{align}
    \label{eq:loss_training}
    \mathcal{L}_{\vtheta, \vphi} = \bm{1}\tran \, \big[ & \bm{\lambda}_\vtheta(\tilde\vz) \odot |\vh_Q - \mq\vmu_\vtheta(\tilde\vz)|^2 - \log\bm{\lambda}_\vtheta(\tilde\vz)  \\
    & -\log\vsig_\vphi(\vy) + 0.5(\vmu^2_\vphi(\vy) + \,\vsig^2_\vphi(\vy)) \big]. \nonumber
\end{align}
The argument and constants are omitted for brevity and $\tilde\vz$ is a sample from $q_\vphi(\vz\cnd\vy)$.
Since VAE-genie has the ground-truth channel as encoder input, the training loss for this variant replaces $\vmu_\vphi(\vy)$ and $\vsig_\vphi(\vy)$ with $\vmu_\vphi(\vh)$ and $\vsig_\vphi(\vh)$, respectively. 
Apart from that, the training loss is identical to~\eqref{eq:loss_training}.
For the training of VAE-real, in~\eqref{eq:loss_training}, $\vh_Q$ is replaced with $\vy_Q=\mq\ma\herm\vy$, and $\bm{\lambda}_\vtheta(\vz)$ with $(\vc_\vtheta(\vz)+\varsigma^2\bm{1})^{-1}$.
Table~\ref{tab:vae-variants} summarizes the proposed estimator variants with an overview regarding the respective encoder input and training loss. 
In each case, the training loss refers to a single batch element.

\begin{table*}[t]
    \caption{Overview of the proposed \ac{vae}-based estimator variants.}
    \centering
    \renewcommand{\arraystretch}{1.5}
    \label{tab:vae-variants}
    \begin{tabular}{l|c|c}
        \hline
        variant & \ac{vae} encoder input & training loss $\mathcal{L}_{\vtheta,\vphi}$ (relates to one batch element) \\ \hline
        VAE-genie & ground-truth channel & $\bm{1}\tran  \big[ \bm{\lambda}_\vtheta(\tilde\vz) \odot |\vh_Q - \mq\vmu_\vtheta(\tilde\vz)|^2 - \log\bm{\lambda}_\vtheta(\tilde\vz) - \log\vsig_\vphi(\vh) + 0.5\,(\vmu^2_\vphi(\vh) + \,\vsig^2_\vphi(\vh)) \big]$ \\ 
        VAE-noisy & noisy observation & $\bm{1}\tran  \big[ \bm{\lambda}_\vtheta(\tilde\vz) \odot |\vh_Q - \mq\vmu_\vtheta(\tilde\vz)|^2 - \log\bm{\lambda}_\vtheta(\tilde\vz) - \log\vsig_\vphi(\vy) + 0.5\,(\vmu^2_\vphi(\vy) + \,\vsig^2_\vphi(\vy)) \big]$ \\ 
        VAE-real & noisy observation & $\bm{1}\tran  \big[ (\vc_\vtheta(\tilde\vz)+\varsigma^2\bm{1})^{-1} \odot |\vy_Q - \mq\vmu_\vtheta(\tilde\vz)|^2 + \log(\vc_\vtheta(\tilde\vz) + \varsigma^2\bm{1}) - \log\vsig_\vphi(\vy) + 0.5\,(\vmu^2_\vphi(\vy) + \,\vsig^2_\vphi(\vy)) \big]$ \\ \hline
    \end{tabular}
\end{table*}

\begin{figure*}[t!]
    \centering
    \def\arrowscale{0.02}
    \input{figures/vae_layers}
    \caption{Detailed illustration of the different layers constituting our VAE implementation. The real and imaginary parts of the input are stacked as \acfp{cc} and processed. The colored arrows represent different layers or layer compositions. Purple \includegraphics{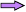} stands for a $1\times 1$ \ac{cl}, orange \includegraphics{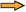} for a block of a \acf{cl}, \acf{bn} layer, and ReLU activation function, gray \includegraphics{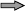} for a \acf{rl}, green \includegraphics{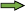} for a \acf{ll}, and red \includegraphics{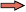} for a block of a transposed \ac{cl}, \ac{bn} layer, and ReLU activation function.}
    \label{fig:vae_layers}
\end{figure*}

We briefly describe our \ac{vae} implementation at this point. 
The simulation code with the corresponding architectures is also publicly available.\footnote{\url{https://github.com/tum-msv/vae-estimator}.}
Fig.~\ref{fig:vae_layers} illustrates the \ac{vae} implementation. 
The arrow colorings in Fig.~\ref{fig:vae_layers} symbolize different layers or layer compositions.
On the left, it is visible that the real and imaginary parts of the encoder input are stacked as \acp{cc}. 
As a first block, the purple arrow \includegraphics[scale=\arrowscale]{figures/arrows_purple.pdf} represents a $1\times 1$ \ac{cl} that maps to a higher number of \acp{cc}, which is different for every system configuration. 
Subsequently, three orange arrows \includegraphics[scale=\arrowscale]{figures/arrows_orange.pdf} follow, representing a block of a \ac{cl}, a \ac{bn} layer, and a ReLU activation function. 
In each \ac{cl}, the \ac{cc} amount at the output is multiplied by a factor of $1.75$.
After a \ac{rl} and \ac{ll}, symbolized by the gray arrow \includegraphics[scale=\arrowscale]{figures/arrows_gray.pdf} and green arrow \includegraphics[scale=\arrowscale]{figures/arrows_green.pdf}, respectively, we arrive at the latent space. 
The reparameterized sample $\vz$ is fed into the decoder, which is a symmetrically flipped version of the encoder. 
The red arrows \includegraphics[scale=\arrowscale]{figures/arrows_red.pdf} symbolize blocks of a transposed \ac{cl}, a \ac{bn} layer, and a ReLU activation function. 
At the output, we have a sample with three \acp{cc} that is fed into an \ac{rl} and \ac{ll} to produce the decoder outputs. 
We use exponential functions to enforce strictly positive values as it is required for $\vsig_\vphi$ and $\vc_\vtheta$.

The number of \acp{cc}, kernel size, and latent dimension are different for every system configuration and are found by a random search over the hyperparameter space by searching for the combination that yields the highest value for~\eqref{eq:dec-like-diag}~\cite{Bergstra2012}. 
We perform the random search with the help of the \textit{Tune} package~\cite{Liaw2018}. 
We use 2D \acp{cl} in the \ac{mimo} case and 1D \acp{cl} in the \ac{simo} case. 
A batch size of $128$, a learning rate of $7\cdot 10^{-4}$ in combination with Adam~\cite{Kingma2019}, and a stride of two in the \acp{cl} are used. 
Overall, we found the estimation performance is robust regarding the selected \ac{vae} architecture as long as it contains enough layers and model parameters for the considered problem.
We implement the \acp{dnn} with PyTorch and refer the reader to the simulation code for further details. 
Additionally, we experimented with \ac{bn} and its variants to determine how we can achieve the best performance~\cite{Ioffe2015, Salimans2016a, Ba2016}. 
We achieve the best performance with \ac{bn} as is proposed in~\cite{Ioffe2015}. 
The only important point is to consider a large enough batch size to limit the variance of the stochastic gradient. 
We additionally use the method of free bits during the training as described in~\cite{Kingma2019}.


\subsection{Computational Complexity}
\label{subsec:complexity}

In this section, we discuss the computational complexity of the proposed estimator. 
The procedure to determine $\hat\vh_{\text{VAE}}(\vy)$ can be split into two parts.
The first step is a forward pass through the \ac{vae} to acquire $\vmu_\vtheta(\vz)$ and $\mc_\vtheta(\vz)$.
The second step is the evaluation of $t_\vtheta(\vz, \vy)$ in~\eqref{eq:lmmse-vae} with given $\vmu_\vtheta(\vz)$ and $\mc_\vtheta(\vz)$.
The computational complexity of the first step is tied to the \ac{vae} architecture in Fig.~\ref{fig:vae_layers}. 
Since all layers exhibit a different complexity, we need a complexity bound for which two aspects are relevant. 
First, a \ac{cl} requires $\OO(R N)$ time, with $R$ being the product of the number of parameters divided by the stride in the \ac{cl}. 
Second, the final \ac{ll} requires $\OO(N^2)$ time.
The remaining layers exhibit less complexity than the \acp{cl} and final \ac{ll}. 
Although $R$ should be increased if $N$ grows, $R$ arguably does not show more than linear growth in $N$.
In conclusion, utilizing $\OO(N^2)$ as complexity bound per layer is reasonable. 
For the $D$ layers of the \ac{vae} forward pass, this makes an overall complexity of $\OO(D N^2)$.

We come to the second step of obtaining $\hat\vh_{\text{VAE}}(\vy)$, which is the evaluation of $t_\vtheta(\vz, \vy)$.
In principle, the inversion of $\ma\mc_\vtheta(\vz)\ma\herm+\varsigma^2\mathbf{I}$ dominates the complexity.
Let us inspect $\ma\mc_\vtheta(\vz)\ma\herm = \ma\,\mq\herm \diag(\vc_\vtheta(\vz)) \mq\,\ma\herm$ in more detail.
If we assume to have unitary pilots and set $\tilde\ma = \ma\mq\herm$ we can show that $\tilde\ma\tilde\ma\herm=\mathbf{I}$ and $\tilde\ma\herm\tilde\ma=\mathbf{I}$ holds, so $\tilde\ma$ is unitary. 
Hence, the inverse of $\ma\mc_\vtheta(\vz)\ma\herm$ is $\tilde\ma\diag(\vc^{-1}_\vtheta(\vz))\tilde\ma\herm$.
We can therefore simplify the estimate $t_\vtheta(\vz, \vy)$ as in~\eqref{eq:lmmse-vae} to
\begin{equation}
    \vmu_\vtheta(\vz) + \mq\herm \diag(\bm{1} + \vc_\vtheta(\vz) \odot \varsigma^{-2}\bm{1}) \mq (\ma\herm\vy - \vmu_\vtheta(\vz))
    \label{eq:lmmse-vae-simple}
\end{equation}
whose complexity is $\OO(N\log N)$ due to multiplying with $\mq$.
As can be seen from our elaborations above, the evaluation of the \ac{vae} requires $\OO(D N^2)$ time, which outweighs the evaluation time of $\OO(N \log N)$ for~\eqref{eq:lmmse-vae-simple}. 
Additionally, many potentials exist to reduce the \ac{vae} complexity, e.g., with pruning~\cite{Anwar2017}. 
Moreover, the computations in the \ac{vae} are highly parallelizable due to the \acp{cl}, which mitigates the $\OO(D N^2)$ complexity.

\subsection{Channel Models}
\label{subsec:channel-models}

We consider different channel models in this work to validate the proposed methods. 
The \ac{3gpp} defines an urban macrocell spatial channel model which computes the transmit-side \ac{ccm} as~\cite{3GPP2020}:
\begin{equation}
    \mc_{\vdel,\text{tx}}=\int_{-\pi}^{\pi} g_{\text{tx}}(\vartheta;\vdel) \va_{\text{tx}}(\vartheta) \va^{\mathrm H}_{\text{tx}}(\vartheta) \diff \vartheta.
    \label{eq:ccm-3gpp}
\end{equation}
The vector $\va_\text{tx}(\vartheta)$ denotes the transmit array steering vector, which is $[1, \exp(\jim \pi \sin(\vartheta)), \ldots, \exp(\jim \pi (\Ntx-1) \sin(\vartheta))]\herm$ in the case of a \ac{ula}.
Analogously, the receive-side \ac{ccm} $\mc_{\vdel,\text{rx}}$ is obtained.
The function $g_\text{tx}(\cdot;\vdel)$ describes an angular power spectrum relating to the involved propagation clusters and is parameterized by the vector $\vdel$, following a prior distribution $p(\vdel)$ that accounts for the involved path gains and angles. 
More precisely, $g_\text{tx}(\cdot;\vdel)$ is a mixture of Laplace densities whose standard deviations represent the angular spreads, cf.~\cite{Neumann2018} for more details. 
The \ac{ccm} for $\bm h$ in~\eqref{eq:system-inv_prob} is determined as $\mc_\vdel=(\mc_{\vdel,\text{tx}} \otimes \mc_{\vdel,\text{rx}})$, under the assumption of uncorrelated scattering~\cite{Kermoal2002}.
For a large number of sub-paths per propagation cluster, which is a common assumption for sub-\SI{6}{\giga\hertz} frequency bands, a \ac{cg} channel distribution is well-motivated by the central limit theorem.
A channel realization can, thus, be obtained with $\vh \mid \vdel \sim \mathcal{N}_{\CC}(\bm{0}, \mc_{\vdel})$~\cite[Sec. 4.2]{Tse2005}.
Accordingly, a correlated Rayleigh fading model is enforced that only holds conditionally, meaning that every channel is individually associated with a different set of path gains and angles contained in~$\vdel$ representing different propagation clusters. 
Note that $\mc_{\vdel}$ is different for every channel realization, causing $p(\vh)$ to be non-Gaussian.

The \quadriga channel simulator allows for the simulation of realistic channels with spatial consistency and time evolution~\cite{Jaeckel2014, QUADRIGA2016}. 
\ac{mimo} channel matrices are modeled as a superposition of in total $L$ propagation paths such that $\mh = \sum_{\ell=1}^L \mg_\ell \exp(-2\pi \jim f_c \tau_\ell)$ where the carrier frequency is denoted as $f_c$ and the delay of the $\ell$-th path as $\tau_\ell$. 
The entries of the matrix $\mg_\ell$ represent the complex-valued gain between every antenna pair caused by the path loss, antenna radiation pattern, and polarization. 
We use version $2.6.1$ of \quadriga to simulate channels at a frequency of \SI{6}{\giga\hertz} in an urban macrocell scenario. 
We place the \ac{bs} at a height of \SI{25}{\meter}, and it covers a sector of \SI{120}{\degree}.
Twenty percent of the \acp{mt} are outdoors \SI{1.5}{\meter} above the ground at a distance between $35$ and \SI{500}{\meter} from the \ac{bs}. 
The remaining eighty percent are situated indoors at different floor levels. 
We consider a \ac{los} propagation environment, where $L=37$.
We equip the \ac{bs} with ``3GPP-3D'' antennas and the \acp{mt} with omnidirectional antennas. 
After generation, the channels are post-processed to normalize the path gain. 
Compared to the \ac{3gpp} channel model, which is fully stochastic, the \quadriga simulator is of a geometric nature.
\quadriga determines channel realizations by a geometric simulation in a randomized and approximately realistic \ac{bs} environment. 
The \quadriga model enables us to highlight that the proposed framework works independently of the adopted channel model.

\subsection{Related Channel Estimators}
\label{subsec:related-est}

This section presents related channel estimators as baselines for the numerical simulations in Section~\ref{subsec:res_nmse}. 
In the case of the 3GPP channel model from Section~\ref{subsec:channel-models}, we have access to the true \ac{ccm} $\mc_{\vdel}$. 
This allows us to evaluate a genie covariance-based estimator (genie-cov)~\cite{Neumann2018}, which is given by the \ac{lmmse} formula
\begin{equation}
    \hat{\vh}_{\text{genie-cov}}(\vy) = \mc_{\vdel} \ma\herm (\ma\mc_{\vdel}\ma\herm + \msig)^{-1} \vy.
    \label{eq:genie-cov}
\end{equation}
This estimator uses utopian genie knowledge to acquire $\mc_{\vdel}$.

A practical estimator can be based on the sample covariance matrix $\hat{\mc}= \frac{1}{T_{\text{r}}} \sum_{i=1}^{T_{\text{r}}} \vh_i\vh^{\mathrm H}_i$ for $T_{\text{r}}$ samples in the training dataset.
The corresponding estimator reads as 
\begin{equation}
    \hat{\vh}_{\text{global-cov}}(\vy) = \hat\mc \ma\herm (\ma\hat\mc\ma\herm + \msig)^{-1} \vy.
    \label{eq:global-cov}
\end{equation}

\Ac{ls} estimation is another comparison method we investigate in our simulations.
An \ac{ls} estimate can be obtained as $\hat{\vh}_{\text{LS}}(\vy) = \ma\herm\vy$.

\Ac{cs}-based \ac{ce} techniques are another prominent topic in the literature.
Especially regarding millimeter waves, \ac{cs} algorithms are potentially interesting candidates~\cite{Busari2018}. 
This work compares the proposed estimators with the \ac{amp} algorithm~\cite{Donoho2010,Maleki2013}. 
As a dictionary for \ac{amp}, we use a two times oversampled \ac{dft} matrix.

We also want to compare the proposed estimators with current \ac{ml}-based channel estimators. 
A recently proposed method exploits structural information of the MMSE estimator to design a neural network-based estimator for the \ac{simo} signal model~\cite{Neumann2018}. 
The derivation leads to a convolutional neural network with ReLU activation function, so we call this estimator CNN. 
The extension of~\cite{Neumann2018} to the \ac{mimo} case is proposed in~\cite{Fesl2021}, to which we also refer in our simulations.

The last comparison method in this section, also recently proposed, is based on a \ac{gmm}~\cite{Koller2022}. 
The idea is to fit a \ac{gmm} to the underlying channel distribution and parameterize a channel estimator with the help of the \ac{gmm}, representing an estimator based on a generative prior. 
We fit a \ac{gmm} with $128$ mixture components for all simulations and a restriction on the fitted covariances such that they are block-circulant.

\section{Simulation Results}
\label{sec:results}

\pgfplotsset{every axis plot/.append style={very thick}}
\pgfplotsset{every axis/.append style={font=\small}}

This section presents the \ac{ce} results based on numerical simulations.
We create $200{,}000$ channel realizations for every system configuration in the upcoming section representing a randomly sampled realistic \ac{bs} environment.
The channels are divided into $T_{\text{r}}=180{,}000$ training, $T_{\text{v}}=10{,}000$ validation, and $T_{\text{e}}=10{,}000$ test samples.
The channels are normalized such that $\E[\|\vh\|^2]=N$.
In our experiments, we calculate the \ac{nmse} as $\frac{1}{T_{\text{e}}N} \sum_{i=1}^{T_{\text{e}}}\|\vh_i-\hat{\vh}_i\|^2$ for the test dataset, where we denote the $i$-th channel realization and corresponding estimate as $\vh_i$ and $\hat\vh_i$, respectively.
Accordingly, we define the \ac{snr} as $\Ntx/\varsigma^2$.
We train the \acp{vae} for a range of \ac{snr} values between $-19$ and \SI{39}{\decibel}.
The proposed estimators are, therefore, SNR-independent.
During the training of VAE-noisy and VAE-real, we sample new realizations $\vn$ after every epoch.
We train the \acp{vae} until~\eqref{eq:dec-like-diag} does not improve for $100$ consecutive epochs on the validation dataset.
If not stated otherwise, $\NL=16$ for one propagation cluster and $\NL=32$ in all other cases.

\subsection{Numerical Convergence Analysis}
\label{subsec:res_arcstudy}

\begin{figure}[!t]
    \centering
    \begin{tikzpicture}
    \def\pltwww{170pt}
    \def\plthhh{60pt}
    \pgfplotsset{set layers}
	\begin{axis}
		[   
        scale only axis,
		xlabel={training epochs},
		ylabel={loss},
        axis y line*=left,
		xmin=1, xmax=170,
        ymin=10, ymax=1200,
		ymode=log,
        xtick={20,40,60,80,100,120,140,160},
		width=\pltwww,
		height=\plthhh,
		legend pos=north east,
		legend style={font=\scriptsize},
		legend columns=3,
		legend style={at={(0,1)},anchor=north west}
		] 
		\addplot[TUMBeamerBlue, line width=\lineWidth]
		table [ignore chars=", x=ep, y=elbo_m, col sep=comma]{data/loss-vae_circ-ThreeGPP-vae_loss_plot_3p.txt};
		\addlegendentry{ELBO}
  
		\addplot[TUMBeamerGreen, line width=\lineWidth]
		table [ignore chars=", x=ep, y=kl_m, col sep=comma]{data/loss-vae_circ-ThreeGPP-vae_loss_plot_3p.txt};
		\addlegendentry{KL}
  
		\addplot[TUMBeamerRed, line width=\lineWidth]
		table [ignore chars=", x=ep, y=rec_m, col sep=comma]{data/loss-vae_circ-ThreeGPP-vae_loss_plot_3p.txt};
		\addlegendentry{REC}  
  
		
	\end{axis}

    \begin{axis}
		[   
        scale only axis,
		xlabel={epochs},
		ylabel={normalized MSE},
        axis x line=none,
        axis y line*=right,
		xmin=1, xmax=170,
        ymin=0.01, ymax=0.11,
        ymode=log,
		width=\pltwww,
		height=\plthhh,
		legend pos=north east,
		legend style={font=\scriptsize},
		legend columns=1,
		legend style={at={(1,1)},anchor=north east}
		] 
        \addplot[TUMBeamerOrange, line width=\lineWidth]
		table [ignore chars=", x=ep, y=mse_m, col sep=comma]{data/mse-vae_circ-ThreeGPP-vae_loss_plot_3p.txt};
		\addlegendentry{NMSE}  
        
    \end{axis}
\end{tikzpicture}
    \vspace{-3mm}
    \caption{Training of the VAE-noisy variant for the \ac{3gpp} channel model (\ac{simo} case) with three propagation clusters at an \ac{snr} of \SI{10}{\decibel}. ELBO refers to the complete training loss in~\eqref{eq:loss_training}, REC to the negative of~\eqref{eq:dec-like-diag}, and KL to~\eqref{eq:vae-kl}. REC is plotted including the in~\eqref{eq:dec-like-diag} omitted constants.
    }
    \label{fig:elbo_epochs}
\end{figure}
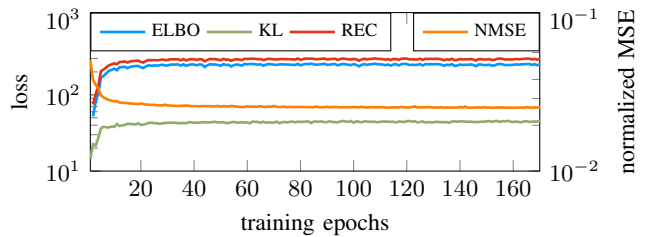

At first, we illustrate the training progress of the VAE-noisy variant for the \ac{3gpp} \ac{simo} signal model with three propagation clusters at an \ac{snr} of \SI{10}{\decibel} in Fig.~\ref{fig:elbo_epochs}. 
The VAE-genie and VAE-real variants exhibit a similar behavior, so we only display VAE-noisy here. 
It can be observed that most of the training progress happens in the first $20$ epochs. 
Interestingly, an increase of the REC term from~\eqref{eq:dec-like-diag} coincides with a decrease of the \ac{nmse}, which indicates that the \ac{vae} learns to properly model the data. 
Moreover, this validates the theoretical analysis in Theorem~\ref{cor:cme_bound} that showed a smaller gap to the \ac{cme} and, thus, a lower \ac{nmse} for a \ac{vae} that better matches the first moments of the posterior distributions, which is achieved during a successful training.

\begin{figure}[!t]
    \centering
    \begin{tikzpicture}
	\begin{axis}
		[   
		xlabel={training samples},
		ylabel={normalized MSE},
		xmin=100, xmax=100000,
		xmode=log,
		ymajorgrids=true,
		xmajorgrids=true,
		grid style=solid,
		grid=both,
		legend pos=north east,
		width=\pltw,
		height=\plthsmall,
		legend style={font=\scriptsize},
		legend columns=1,
		legend style={at={(1,1)},anchor=north east}
		] 
		\addplot[VAEgenie]
		table [ignore chars=", x=samples, y=mse, col sep=comma]{data/results-mse-3gpp-3p-128rx-vae_genie-samples-10db.txt};
		\addlegendentry{\legVAEgenie}
		
		\addplot[VAEnoisy]
		table [ignore chars=", x=samples, y=mse, col sep=comma]{data/results-mse-3gpp-3p-128rx-vae_noisy-samples-10db.txt};
		\addlegendentry{\legVAEnoisy}
		
		\addplot[VAEreal]
		table [ignore chars=", x=samples, y=mse, col sep=comma]{data/results-mse-3gpp-3p-128rx-vae_real-samples-10db.txt};
		\addlegendentry{\legVAEreal}    			
		
		\addplot[geniecov]
		table {%
			10 2.292354777455329895e-02
			100 2.292354777455329895e-02
			1000 2.292354777455329895e-02
			10000 2.292354777455329895e-02
			100000 2.292354777455329895e-02
		};
		\addlegendentry{\leggeniecov}
		
		\addplot[VAEgenie, style=dotted]
		table {%
			10 2.259193360805511475e-02
			100000 2.259193360805511475e-02
		};
		
		\addplot[VAEnoisy, style=dotted]
		table {%
			10 2.499481476843357086e-02
			100000 2.499481476843357086e-02
		};
		
		\addplot[VAEreal, style=dotted]
		table {%
			10 2.543632872402667999e-02
			100000 2.543632872402667999e-02
		};
		
	\end{axis}
\end{tikzpicture}
    \vspace{-3mm}
    \caption{Normalized \ac{mse} for different numbers of training samples at an SNR of \SI{10}{\decibel} for the \ac{3gpp} channel model (\ac{simo} case) with three propagation clusters and $128$ antennas at the receiver. The dotted lines display the achieved result with the complete training dataset of $180{,}000$ samples.}
    \label{fig:training_samples}
\end{figure}
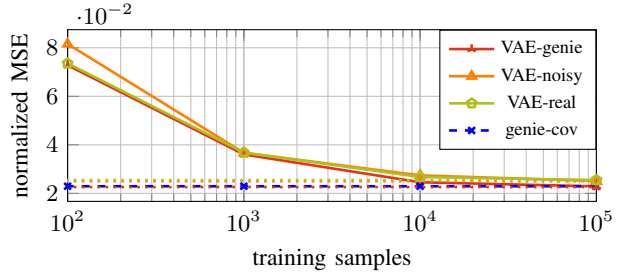

Further, we investigate two critical quantities of the model selection process: the training dataset's size and the latent space's dimensionality. 
Regarding the size of the training dataset, a larger size is likely to lead to better estimation results.
Fig.~\ref{fig:training_samples} provides insights into this matter. 
We display the estimation results of the test dataset for the three proposed variants of \ac{vae}-based estimators depending on the size of the training dataset. 
The \ac{3gpp} channel model (\ac{simo} case) with three propagation clusters and $128$ antennas at the receiver is used in Fig.~\ref{fig:training_samples}. 
As dotted lines, we display the attained estimation result for the complete training dataset of $180{,}000$ samples.
We also show the estimation performance of the genie-cov estimator in blue. 
It is visible that the most progress is reached from $10^2$ to $10^4$ training samples. 
More than $10^4$ training samples only lead to minor \ac{nmse} improvements for all three types of \ac{vae}-based estimators.

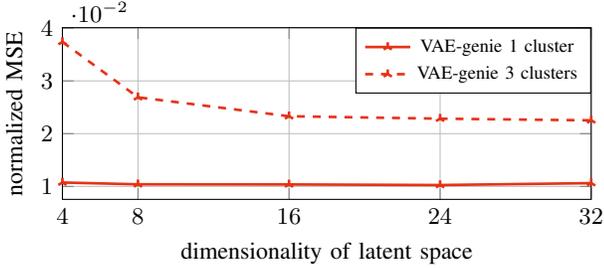
\begin{figure}[!t]
    \centering
    \begin{tikzpicture}
	\begin{axis}
		[   
		xlabel={latent space dimensionality},
		ylabel={normalized MSE},
		xmin=4, xmax=32,
		xtick={4,8,16,24,32},
		ymajorgrids=true,
		xmajorgrids=true,
		grid style=solid,
		grid=both,
		legend pos=north east,
		width=\pltw,
		height=\plthsmall,
		legend style={font=\scriptsize},
		legend columns=1,
		legend style={at={(1,1)},anchor=north east}
		] 
		\addplot[VAEgenie]
		table [ignore chars=", x=lat, y=mse, col sep=comma]{data/results-mse-3gpp-1p-128rx-latent-10db-vae_genie.txt};
		\addlegendentry{VAE-genie 1 cluster}
		
		\addplot[VAEgenie, style=dashed]
		table [ignore chars=", x=lat, y=mse, col sep=comma]{data/results-mse-3gpp-3p-128rx-latent-10db-vae_genie.txt};
		\addlegendentry{VAE-genie 3 clusters}
		
	\end{axis}
\end{tikzpicture}
    \vspace{-3mm}
    \caption{Normalized \ac{mse} for different sizes of the latent space at an SNR of \SI{10}{\decibel} for the \ac{3gpp} channel model (\ac{simo} case) with one or three propagation clusters and $128$ antennas at the receiver.}
    \label{fig:latent_size}
\end{figure}

The influence of the dimensionality of the latent space on the estimation result is less apparent than the size of the training dataset. 
We illustrate the \ac{nmse} for dimensionalities in the range $[4,32]$ for the \ac{3gpp} \ac{simo} channel model with $128$ antennas at the receiver in Fig.~\ref{fig:latent_size} by considering one and three propagation clusters. 
The \ac{nmse} is nearly constant for the case with one propagation cluster. 
In contrast, the \ac{nmse} decreases from dimensionality $4$ to $16$ for the three propagation clusters case and saturates for larger dimensional latent spaces. 
In practice, the operator must select an ample enough latent space to obtain a desirable performance.

\begin{figure}[!t]
    \begin{tikzpicture}
    \def\pltwww{165pt}
    \def\plthhh{60pt}
    \pgfplotsset{set layers}
    \begin{axis}
        [   
        scale only axis,
        xlabel={training epochs},
        ylabel={$\tr(\diag\vsig_\vphi^2)$},
        axis y line*=left,
        xmin=1, xmax=170,
        ymin=0.004, ymax=10,
        ymode=log,
        xtick={20,40,60,80,100,120,140,160},
		width=\pltwww,
		height=\plthhh,
        legend pos=north east,
        legend style={font=\scriptsize},
        legend columns=2,
        legend style={at={(1,1)},anchor=north east}
        ] 
        \addplot[color=TUMBeamerRed, line width=\lineWidth]
        table [ignore chars=", x=ep, y=enc_var, col sep=comma]{data/enc_var-vae_circ-ThreeGPP-enc_var-genie-128rx-1p-lat-4.txt};
        \addlegendentry{VAE-genie}
  
        \addplot[color=TUMBeamerOrange, line width=\lineWidth]
        table [ignore chars=", x=ep, y=enc_var, col sep=comma]{data/enc_var-vae_circ-ThreeGPP-enc_var-noisy-128rx-1p-lat-4.txt};
        \addlegendentry{VAE-noisy}
        
    \end{axis}

    \begin{axis}
        [   
        scale only axis,
        xlabel={epochs},
        ylabel={normalized MSE},
        axis x line=none,
        axis y line*=right,
        xmin=1, xmax=170,
        ymin=0.01, ymax=0.1,
        ymode=log,
		width=\pltwww,
		height=\plthhh,
        legend pos=north east,
        legend style={font=\scriptsize},
        legend columns=4,
        legend style={at={(1,1)},anchor=south east}
        ] 
        \addplot[color=TUMBeamerRed, line width=\lineWidth, dashed]
        table [ignore chars=", x=ep, y=mse_m, col sep=comma]{data/mse-vae_circ-ThreeGPP-enc_var-genie-128rx-1p-lat-4.txt};
  
        \addplot[color=TUMBeamerOrange, line width=\lineWidth, dashed]
        table [ignore chars=", x=ep, y=mse_m, col sep=comma]{data/mse-vae_circ-ThreeGPP-enc_var-noisy-128rx-1p-lat-4.txt};
        
    \end{axis}
\end{tikzpicture}
    \vspace{-3mm}
    \caption{Trace of encoder variances and \ac{nmse} over training epochs on the \ac{3gpp} channel model (\ac{simo} case) with $128$ antennas, one propagation cluster, \SI{10}{\decibel} \ac{snr}, and $\NL=4$. The \ac{nmse} curves are displayed as dashed.}
    \label{fig:enc_var}
\end{figure}
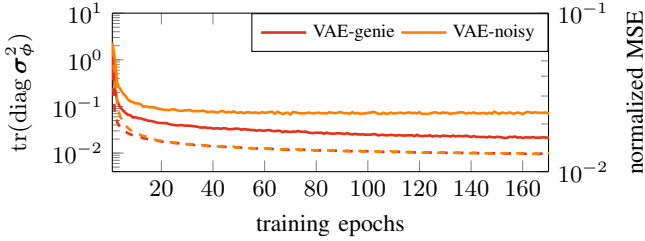

Theorem~\ref{cor:cme_bound} in Section~\ref{subsec:convergence} described that the convergence of the \mapvae estimator to the \ac{cme} depends on the vanishing of $\tr(\mc_{p_{\vtheta}(\vz\cnd\vy)})$.
To this end, we analyze the trace of the encoder variance, i.e., the summed variance $\vsig_\vphi^2$ of $q_\vphi$, which optimally is a good approximation of $\tr(\mc_{p_{\vtheta}(\vz\cnd\vy)})$, over the training epochs on the validation dataset in Fig.~\ref{fig:enc_var}.  
We evaluate VAE-genie and VAE-noisy with $\NL=4$ on the \ac{3gpp} channel model (\ac{simo} case) with $128$ antennas and one propagation cluster at an \ac{snr} of \SI{10}{\decibel}. 
It is visible that both encoder variances are decreasing in a comparable way as the \ac{nmse}, indicating that lower traces improve the \ac{nmse}. 
The noise variance detrimentally influences VAE-noisy's encoder variance trace since it is always higher than that of VAE-genie. 
In conclusion, since the wireless channel data is expected to be sparse or compressible in the angular domain, cf. Section~\ref{sec:practical}, the \ac{vae} indeed aims to find a less stochastic encoder mapping during training. 
This is in agreement with the argumentation in Section~\ref{subsec:convergence} and the observation of a decreasing \ac{nmse} of the parameterized estimator during training in Fig.~\ref{fig:elbo_epochs}.

\begin{figure}[!t]
    \centering
    \begin{tikzpicture}
	\begin{semilogxaxis}[
		xlabel={number of latent samples $K$},
		ylabel={normalized MSE},
		xmin=1, xmax=128,
        xticklabels={1,2,4,8,16,32,64,128},
        log basis x={2},
		ymajorgrids=true,
		xmajorgrids=true,
		grid style=solid,
		grid=both,
		legend pos=north east,
		width=\pltw,
		height=\plthsmall,
		legend style={font=\scriptsize},
		legend style={at={(1,1)},anchor=north east},
		legend columns=3
	] 
	\addplot[VAEgenie]
		table [ignore chars=", x=samples, y=mse, col sep=comma]{data/results-samples-3gpp-3p-128rx-10dB-mc-vae_genie.txt};
	\addlegendentry{\legVAEgenie}	
	
	\addplot[VAEnoisy]
		table [ignore chars=", x=samples, y=mse, col sep=comma]{data/results-samples-3gpp-3p-128rx-10dB-mc-vae_noisy.txt};
	\addlegendentry{\legVAEnoisy}
	
	\addplot[VAEreal]
		table [ignore chars=", x=samples, y=mse, col sep=comma]{data/results-samples-3gpp-3p-128rx-10dB-mc-vae_real.txt};
	\addlegendentry{\legVAEreal}
	
	\addplot[VAEgenie, style=dashed]
		table {%
		1 0.0228993718
		128 0.0228993718
		};

	\addplot[VAEnoisy, style=dashed]
		table {%
		1 0.0253674032
		128 0.0253674032
		};

	\addplot[VAEreal, style=dashed]
		table {%
		1 0.0259267
		128 0.0259267
		};

	\end{semilogxaxis}
\end{tikzpicture}
    \vspace{-3mm}
    \caption{Normalized \ac{mse} for different numbers of samples $K$ drawn in the latent space for the evaluation of $\hat\vh_{\text{VAE}}^{(K)}(\vy)$ at an SNR of \SI{10}{\decibel} for the \ac{3gpp} channel model (\ac{simo} case) with three propagation clusters and $128$ antennas at the receiver. The dashed lines represent the estimate $\hat\vh_\text{VAE}(\vy)$ in~\eqref{eq:vae-estimator-mu}.}
    \label{fig:latent_sampling}
\end{figure}
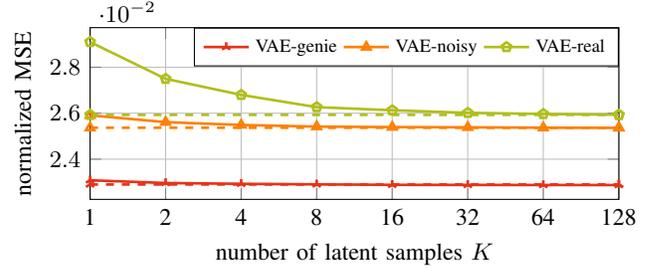

As pointed out in Section~\ref{subsec:mmse-est-vae}, we approximate the \ac{cme} with the \mapvae estimator by only forwarding the latent mean vector $\vmu_\vphi(\vz)$ to approximate the outer expectation in~\eqref{eq:total_exp}.
It is interesting to see the \ac{nmse} performance for different numbers of $K$ samples from $q_\vphi(\vz\cnd\vy)$ to compute $\hat\vh^{(K)}_\text{VAE}(\vy)$ from~\eqref{eq:vae-estimator}.
Fig.~\ref{fig:latent_sampling} provides such an analysis by showing the \ac{nmse} for different numbers of latent samples. 
As dashed lines, we display the \mapvae estimator, which only uses the single sample $\vmu_\vphi(\vz)$ for the input of the \ac{vae}'s decoder. 
We observe that VAE-real benefits the most from more samples. 
For VAE-genie and VAE-noisy, there are only slight improvements present. 
Interestingly, only taking the mean value, representing the \mapvae estimator, delivers an estimation performance of about $K=64$ samples for VAE-real. 
The excellent performance of the \mapvae estimator is a supporting argument for the theoretical analysis of the estimator in Theorem~\ref{cor:cme_bound} that predicts a small distance from the \ac{cme} if the \ac{vae} is well-trained.

\subsection{Normalized MSE Results}
\label{subsec:res_nmse}

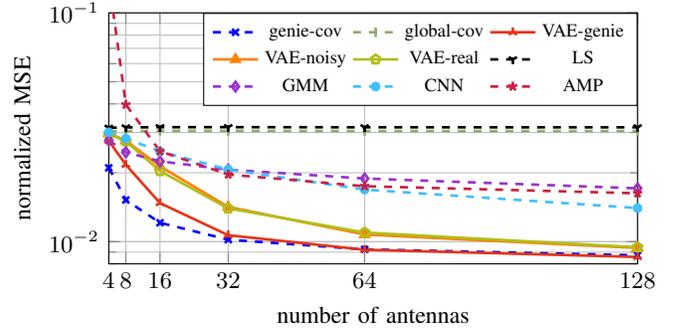
\begin{figure}[!t]
    \centering
    \begin{tikzpicture}
    	\begin{semilogyaxis}[
    		xlabel={number of antennas},
    		ylabel={normalized MSE},
    		xmin=4, xmax=128,
            xtick={4,8,16,32,64,128},
            ymin=8e-3,ymax=1e-1,
    		ymajorgrids=true,
    		xmajorgrids=true,
    		grid=both,
    		grid style=solid,
    		legend pos=south west,
    		width=\pltw,
    		height=\plth,
    		legend style={font=\scriptsize},
    		legend columns=3,
		    legend style={at={(1,1)},anchor=north east}
    	] 
    	\addplot[geniecov]
    		table [ignore chars=", x=ant, y=geniecov, col sep=comma]{data/ant_vs_nmse_3gpp_3p.txt};
    	\addlegendentry{\leggeniecov}
    	
    	\addplot[globalcov]
    		table [ignore chars=", x=ant, y=globalcov, col sep=comma]{data/ant_vs_nmse_3gpp_3p.txt};
    	\addlegendentry{\legglobalcov}
    	
    	\addplot[VAEgenie]
    		table [ignore chars=", x=ant, y=vaegenie, col sep=comma]{data/ant_vs_nmse_3gpp_3p.txt};
    	\addlegendentry{\legVAEgenie}  	
    	
    	\addplot[VAEnoisy]
    		table [ignore chars=", x=ant, y=vaenoisy, col sep=comma]{data/ant_vs_nmse_3gpp_3p.txt};
    	\addlegendentry{\legVAEnoisy}
    	
    	\addplot[VAEreal]
    		table [ignore chars=", x=ant, y=vaereal, col sep=comma]{data/ant_vs_nmse_3gpp_3p.txt};
    	\addlegendentry{\legVAEreal}
    	
    	\addplot[LS]
    		table [ignore chars=", x=ant, y=ls, col sep=comma]{data/ant_vs_nmse_3gpp_3p.txt};
    	\addlegendentry{\legLS}
    	
    	
    	\addplot[GMM circ]
    		table [ignore chars=", x=ant, y=gmmcirc, col sep=comma]{data/ant_vs_nmse_3gpp_3p.txt};
    	\addlegendentry{\leggmmcirc}
    	
    	\addplot[CNN]
    		table [ignore chars=", x=ant, y=cnn, col sep=comma]{data/ant_vs_nmse_3gpp_3p.txt};
    		\addlegendentry{\legcnn}
    		
    	\addplot[AMP]
    		table [ignore chars=", x=ant, y=amp, col sep=comma]{data/ant_vs_nmse_3gpp_3p.txt};
    	\addlegendentry{\legamp}
    	
    	\end{semilogyaxis}
\end{tikzpicture}
    \vspace{-3mm}
    \caption{Normalized \ac{mse} for the \ac{3gpp} channel model (\ac{simo} case) with three propagation clusters for different numbers of antennas at the receiver at an \ac{snr} of \SI{15}{\decibel}. The proposed methods are displayed with solid linestyles.}
    \label{fig:3gpp_ant_3p}
\end{figure}

We begin with an \ac{nmse} investigation for the \ac{3gpp} channel model (\ac{simo} case) with three propagation clusters and varying numbers of antennas at the receiver at an \ac{snr} of \SI{15}{\decibel} in Fig.~\ref{fig:3gpp_ant_3p}. 
The illustration shows that the proposed \ac{vae}-based methods need a sufficiently large amount of antennas to develop their full potential. 
From $16$ antennas on, the \ac{vae}-based methods outperform the baselines and exhibit increasing performance gains if more antennas are considered. 
All other baselines perform significantly worse than the proposed methods in the large antenna regime. 
It is also visible that VAE-genie converges to the genie-cov curve. 
Surprisingly, the VAE-real variant is almost on par with the VAE-noisy variant, although no ground-truth data is available for its training, underlining the strong performance of the \ac{vae} as a generative prior even in cases with imperfect training data.

\begin{figure}[!t]
    \centering
    \begin{tikzpicture}
    	\begin{semilogyaxis}[
    		xlabel={SNR [dB]},
    		ylabel={normalized MSE},
    		xmin=-10, xmax=30,
            xtick={-10,-5,0,5,10,15,20,25,30},
            ymax=1.01, ymin=1.1e-4,
    		ymajorgrids=true,
    		xmajorgrids=true,
    		grid=both,
    		grid style=solid,
    		legend pos=south west,
    		width=\pltw,
    		height=\plth,
    		legend style={font=\scriptsize},
    		legend columns=3,
		    legend style={at={(0,0)},anchor=south west}
    	] 
    	\addplot[geniecov]
    		table [ignore chars=", x=snr, y=mse, col sep=comma]{data/results-mse-3gpp-1p-128rx-genie_cov.txt};
    	\addlegendentry{\leggeniecov}
    	
    	\addplot[VAEgenie]
    		table [ignore chars=", x=snr, y=mse, col sep=comma]{data/results-mse-3gpp-1p-128rx-vae_genie.txt};
    	\addlegendentry{\legVAEgenie}  	
    	
    	\addplot[LS]
    		table [ignore chars=", x=snr, y=mse, col sep=comma]{data/results-mse-3gpp-1p-128rx-ls.txt};
    	\addlegendentry{\legLS}
    	
    	\addplot[globalcov]
    		table [ignore chars=", x=snr, y=mse, col sep=comma]{data/results-mse-3gpp-1p-128rx-global_cov.txt};
    	\addlegendentry{\legglobalcov}
    	
    	\addplot[VAEnoisy]
    		table [ignore chars=", x=snr, y=mse, col sep=comma]{data/results-mse-3gpp-1p-128rx-vae_noisy.txt};
    	\addlegendentry{\legVAEnoisy}
    	
    	\addplot[CNN]
    		table [ignore chars=", x=SNR, y=cnn_fft2x_relu_non_hier_False, col sep=comma]
    		{data/2023-02-13_09-28-41_128antennas_1paths_3gpp=True.csv};
    		\addlegendentry{\legcnn}
    	
    	
    	\addplot[GMM circ]
    		table [ignore chars=", x=SNR, y=gmm, col sep=comma]
    		{data/2022-12-22_09-35-59_n_paths=1_ant=128_comp=128_3gpp=True_diag_simo_baur.csv};
    	\addlegendentry{\leggmmcirc}
    	
    	\addplot[VAEreal]
    		table [ignore chars=", x=snr, y=mse, col sep=comma]{data/results-mse-3gpp-1p-128rx-vae_real.txt};
    	\addlegendentry{\legVAEreal}
    		
    	\addplot[AMP]
    		table [ignore chars=", x=SNR, y=amp_2os, col sep=comma]
    		{data/2023-02-14_12-36-11_n_paths=1_ant=128_comp=_3gpp=True_oversamp=2_simo_baur.csv};
    	\addlegendentry{\legamp}
    	
    	\end{semilogyaxis}
\end{tikzpicture}
    \vspace{-3mm}
    \caption{Normalized \ac{mse} for the \ac{3gpp} channel model (\ac{simo} case) with one propagation cluster and $128$ antennas at the receiver. The proposed methods are displayed with solid linestyles.}
    \label{fig:3gpp_128rx_1p}
\end{figure}
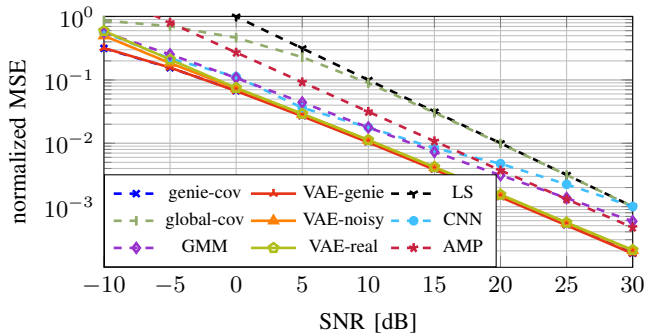

A massive amount of antennas is significant for prospective communications systems. 
Hence, we investigate the large antenna regime in more detail in the following. 
We inspect the \ac{nmse} performance for the \ac{3gpp} channel model with one propagation cluster and $128$ antennas at the receiver in the \ac{simo} case over the \ac{snr} in Fig.~\ref{fig:3gpp_128rx_1p}.
The proposed estimators outperform the baseline methods and achieve about \SI{10}{\decibel} advantage compared to LS over the whole \ac{snr} range.

\begin{figure}[!t]
    \centering
    \begin{tikzpicture}
    	\begin{semilogyaxis}[
    		xlabel={SNR [dB]},
    		ylabel={normalized MSE},
    		xmin=-10, xmax=30,
            ymax=1.01,
            xtick={-10,-5,0,5,10,15,20,25,30},
    		ymajorgrids=true,
    		xmajorgrids=true,
    		grid=both,
    		grid style=solid,
    		legend pos=south west,
    		width=\pltw,
    		height=\plth,
    		legend style={font=\scriptsize},
    		legend columns=2,
    		ymax=1,
		    legend style={at={(0,0)},anchor=south west}
    	] 
    	\addplot[globalcov]
    		table [ignore chars=", x=snr, y=mse, col sep=comma]{data/results-mse-quadriga-los-128rx-global_cov.txt};
    	\addlegendentry{\legglobalcov}
    	
    	\addplot[VAEgenie]
    		table [ignore chars=", x=snr, y=mse, col sep=comma]{data/results-mse-quadriga-los-128rx-vae_genie.txt};
    	\addlegendentry{\legVAEgenie}
    	
    	\addplot[VAEnoisy]
    		table [ignore chars=", x=snr, y=mse, col sep=comma]{data/results-mse-quadriga-los-128rx-vae_noisy.txt};
    	\addlegendentry{\legVAEnoisy}
    	
    	\addplot[VAEreal]
    		table [ignore chars=", x=snr, y=mse, col sep=comma]{data/results-mse-quadriga-los-128rx-vae_real.txt};
    	\addlegendentry{\legVAEreal}
    	
    	\addplot[LS]
    		table [ignore chars=", x=snr, y=mse, col sep=comma]{data/results-mse-quadriga-los-128rx-ls.txt};
    	\addlegendentry{\legLS}
    	
    	
    	\addplot[GMM circ]
    		table [ignore chars=", x=SNR, y=gmm, col sep=comma]
    		{data/2022-12-22_09-36-42_n_paths=1_ant=128_comp=128_3gpp=False_diag_simo_baur.csv};
    	\addlegendentry{\leggmmcirc}
    	
    	\addplot[CNN]
    		table [ignore chars=", x=SNR, y=cnn_fft2x_relu_non_hier_False, col sep=comma]
    		{data/2023-02-13_09-41-58_128antennas_1paths_3gpp=False_LOS.csv};
    		\addlegendentry{\legcnn}
    		
    	\addplot[AMP]
    		table [ignore chars=", x=SNR, y=amp_2os, col sep=comma]
    		{data/2023-02-14_16-55-28_n_paths=1_ant=128_comp=_3gpp=False_oversamp=2_LOS_simo_baur.csv};
    	\addlegendentry{\legamp}
    	
    	\end{semilogyaxis}
\end{tikzpicture}
    \vspace{-3mm}
    \caption{Normalized \ac{mse} for the \quadriga channel model (\ac{simo} case) with LOS channels and $128$ antennas at the receiver. The proposed methods are displayed with solid linestyles.}
    \label{fig:quadriga_128rx_los}
\end{figure}
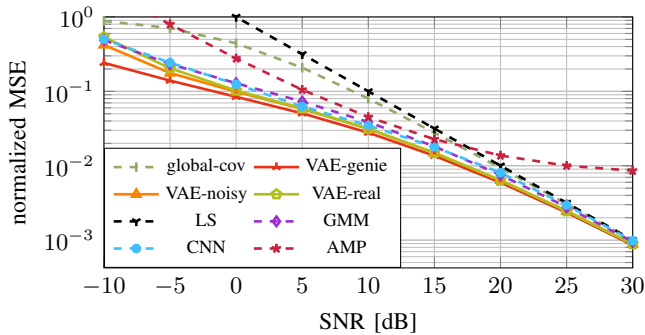

To highlight the independence of the adopted channel model, we show simulation results for the \quadriga channel model (\ac{simo} case) in Fig.~\ref{fig:quadriga_128rx_los}. 
This time, we cannot display the genie-cov curve as the true \ac{ccm} is unavailable. 
As can be seen in the plot, all \ac{vae}-based estimators show superior \ac{nmse} results. 
The advantages are not as pronounced as in the previous figures but still noticeable.

\begin{figure}[!t]
    \centering
    \begin{tikzpicture}
    	\begin{semilogyaxis}[
    		xlabel={SNR [dB]},
    		ylabel={normalized MSE},
    		xmin=-10, xmax=30,
            ymax=1.01, ymin=1.1e-4,
            xtick={-10,-5,0,5,10,15,20,25,30},
    		ymajorgrids=true,
    		xmajorgrids=true,
    		grid=both,
    		grid style=solid,
    		legend pos=south west,
    		width=\pltw,
    		height=\plth,
    		legend style={font=\scriptsize},
    		legend columns=2,
    		ymax=1,
		    legend style={at={(0,0)},anchor=south west}
    	] 
    	\addplot[geniecov]
    		table [ignore chars=", x=snr, y=mse, col sep=comma]{data/results-mse-3gpp-mimo-1p-32rx4tx-genie_cov.txt};
    	\addlegendentry{\leggeniecov}
    	
    	\addplot[globalcov]
    		table [ignore chars=", x=snr, y=mse, col sep=comma]{data/results-mse-3gpp-mimo-1p-32rx4tx-global_cov.txt};
    	\addlegendentry{\legglobalcov}
    	
    	\addplot[VAEgenie]
    		table [ignore chars=", x=snr, y=mse, col sep=comma]{data/results-mse-3gpp-mimo-1p-32rx4tx-vae_genie.txt};
    	\addlegendentry{\legVAEgenie}
    	
    	\addplot[VAEnoisy]
    		table [ignore chars=", x=snr, y=mse, col sep=comma]{data/results-mse-3gpp-mimo-1p-32rx4tx-vae_noisy.txt};
    	\addlegendentry{\legVAEnoisy}
    	
    	\addplot[VAEreal]
    		table [ignore chars=", x=snr, y=mse, col sep=comma]{data/results-mse-3gpp-mimo-1p-32rx4tx-vae_real.txt};
    	\addlegendentry{\legVAEreal}
    	
    	\addplot[LS]
    		table [ignore chars=", x=snr, y=mse, col sep=comma]{data/results-mse-3gpp-mimo-1p-32rx4tx-ls.txt};
    	\addlegendentry{\legLS}
    	
    	
    	\addplot[GMM bcirc]
    		table [ignore chars=", x=SNR, y=GMM_block-diag, col sep=comma]
     		{data/2023-04-28_21-54-53_tx=4_rx=32_train=180000_test=10000_sums=all_comrx=32_comptx=4_comp=128_cov=block-diag_paths=1_3gpp.csv};
    	\addlegendentry{\leggmmbcirc}
    	
    	\addplot[CNN]
    		table [ignore chars=", x=SNR, y=cnn_fft_relu, col sep=comma]
    		{data/2023-02-14_10-30-00_1paths_32BS_4MS_35AS_20epochs_500ebatch_20lbatchsize_4pilots_3gpp.csv};
    	\addlegendentry{\legcnn}
    	
    	\end{semilogyaxis}
\end{tikzpicture}
    \vspace{-3mm}
    \caption{Normalized \ac{mse} for the \ac{3gpp} channel model (\ac{mimo} case) with one propagation cluster, $32$ antennas at the receiver, and $4$ antennas at the transmitter. The proposed methods are displayed with solid linestyles.}
    \label{fig:3gpp_32rx4tx_1p}
\end{figure}

\begin{figure}[!t]
    \centering
    \begin{tikzpicture}
    	\begin{semilogyaxis}[
    		xlabel={SNR [dB]},
    		ylabel={normalized MSE},
    		xmin=-10, xmax=30,
            ymax=1.01, 
            xtick={-10,-5,0,5,10,15,20,25,30},
    		ymajorgrids=true,
    		xmajorgrids=true,
    		grid=both,
    		grid style=solid,
    		legend pos=south west,
    		width=\pltw,
    		height=\plth,
    		legend style={font=\scriptsize},
    		legend columns=2,
    		ymax=1,
		    legend style={at={(0,0)},anchor=south west}
    	] 
    	\addplot[geniecov]
    		table [ignore chars=", x=snr, y=mse, col sep=comma]{data/results-mse-3gpp-mimo-3p-32rx4tx-genie_cov.txt};
    	\addlegendentry{\leggeniecov}
    	
    	\addplot[globalcov]
    		table [ignore chars=", x=snr, y=mse, col sep=comma]{data/results-mse-3gpp-mimo-3p-32rx4tx-global_cov.txt};
    	\addlegendentry{\legglobalcov}
    	
    	\addplot[VAEgenie]
    		table [ignore chars=", x=snr, y=mse, col sep=comma]{data/results-mse-3gpp-mimo-3p-32rx4tx-vae_genie.txt};
    	\addlegendentry{\legVAEgenie}  	
    	
    	\addplot[VAEnoisy]
    		table [ignore chars=", x=snr, y=mse, col sep=comma]{data/results-mse-3gpp-mimo-3p-32rx4tx-vae_noisy.txt};
    	\addlegendentry{\legVAEnoisy}
    	
    	\addplot[VAEreal]
    		table [ignore chars=", x=snr, y=mse, col sep=comma]{data/results-mse-3gpp-mimo-3p-32rx4tx-vae_real.txt};
    	\addlegendentry{\legVAEreal}
    	
    	\addplot[LS]
    		table [ignore chars=", x=snr, y=mse, col sep=comma]{data/results-mse-3gpp-mimo-3p-32rx4tx-ls.txt};
    	\addlegendentry{\legLS}
    	
    	
    	\addplot[GMM bcirc]
    		table [ignore chars=", x=SNR, y=GMM_block-diag, col sep=comma]
    		{data/2023-04-29_11-40-21_tx=4_rx=32_train=180000_test=10000_sums=all_comrx=32_comptx=4_comp=128_cov=block-diag_paths=3_3gpp.csv};
    	\addlegendentry{\leggmmbcirc}
    	
    	\addplot[CNN]
    		table [ignore chars=", x=SNR, y=cnn_fft_relu, col sep=comma]
    		{data/2023-02-14_10-31-20_3paths_32BS_4MS_35AS_20epochs_500ebatch_20lbatchsize_4pilots_3gpp.csv};
    	\addlegendentry{\legcnn}
    	
    	\end{semilogyaxis}
\end{tikzpicture}
    \vspace{-3mm}
    \caption{Normalized \ac{mse} for the \ac{3gpp} channel model (\ac{mimo} case) with three propagation clusters, $32$ antennas at the receiver, and $4$ antennas at the transmitter. The proposed methods are displayed with solid linestyles.}
    \label{fig:3gpp_32rx4tx_3p}
\end{figure}

Fig.~\ref{fig:3gpp_32rx4tx_1p} shows the \ac{nmse} performance for the \ac{3gpp} channel model (\ac{mimo} case) with one propagation cluster, $32$ antennas at the receiver, and $4$ antennas at the transmitter. 
The qualitative behavior of the curves is similar to Fig.~\ref{fig:3gpp_128rx_1p}, where also one propagation cluster is considered.
CNN and GMM show the worst \ac{nmse} among the \ac{ml}-based methods for \acp{snr} larger than \SI{-5}{\decibel}. 
In this case as well, VAE-noisy and VAE-real show comparable performance. 
Compared to LS, the \ac{vae}-based methods attain a performance gain between $6$ and \SI{13}{\decibel}. 
In Fig.~\ref{fig:3gpp_32rx4tx_3p}, we illustrate estimation results for the \ac{3gpp} channel model (\ac{mimo} case) with three propagation clusters, $32$ antennas at the receiver, and $4$ antennas at the transmitter. 
As in the previous figure, the proposed estimators clearly outperform the baselines. 
However, the performance gaps in Fig.~\ref{fig:3gpp_32rx4tx_1p} are more noticeable than in Fig.~\ref{fig:3gpp_32rx4tx_3p}. 
The performance gain compared to LS shrinks to a range from $2$ to \SI{11}{\decibel}.

In summary, the \ac{vae}-based methods exhibit immense performance gains for large antenna arrays, i.e., larger equal $16$ antennas and all considered numbers of propagation clusters, significantly outperforming the baseline methods. 
The strong performance for arrays with many antennas is likely due to the circulant approximation to the Toeplitz \ac{ccm}, which becomes better for large arrays. 
VAE-genie lies almost on the genie-cov estimator, and the performance of VAE-noisy and VAE-real is almost identical for all considered scenarios. 
Moreover, all \ac{ml}-based methods use genie knowledge during the training phase in the form of ground-truth channel training data, except for VAE-real, which is trained and evaluated solely based on noisy pilot observations. 
From this point of view, the strong estimation results of VAE-real are even more meaningful. 
What is more, the parameterization choices for $p_\vtheta(\vh\cnd\vz)$ and $q_\vphi(\vz\cnd\vy)$ excellently fulfill their purpose in accordance with Theorem~\ref{cor:cme_bound} as a result of the \ac{vae}-based approaches' strong estimation results. 
This becomes apparent when comparing the \ac{vae} to the global-cov or \ac{gmm} results, either adopting a Gaussian or \ac{gmm} prior, respectively, highlighting the superiority of the \ac{vae}-based prior.

\section{Conclusion}
\label{sec:conclusion}

This manuscript presents a novel estimation technique based on the \ac{vae}. 
The idea is to tractably model the underlying data distribution as \ac{cg} via a \ac{vae}, yielding a powerful generative prior. 
The \ac{cg} modeling allows us to parameterize the \ac{mse}-optimal \ac{cme} under the \ac{vae} framework. 
We propose three estimator variants, of which we find the VAE-real variant particularly appealing as it does not require access to ground-truth data during training or evaluation. 
We provide theoretical analysis that quantifies the error gap between the proposed \mapvae estimator and the \ac{cme} and relates the training process of the \ac{vae} to the estimation performance, supporting the strong estimation capabilities of the proposed \ac{vae}-based estimators. 
Our extensive \ac{ce} simulations highlight that the proposed methods attain excellent performance for various system configurations. 
In future works, we want to investigate the effects of regularization terms in the training objective and analyze other (especially wide) observation matrices. 
Moreover, we plan to consider nonlinear system models, e.g., quantized systems~\cite{Fesl2024}, to broaden the application horizon.

\appendix
\subsection{Derivation of~\eqref{eq:lmmse-vae}}
\label{app:derivation-lmmse}

For the parameterized joint \ac{pdf} of $\vy$ and $\vh$ given $\vz$, 
\begin{equation}
    p_\vtheta(\vy,\vh\cnd\vz) = p(\vy\cnd\vh,\vz) \, p_\vtheta(\vh\cnd\vz) = p(\vy\cnd\vh) \, p_\vtheta(\vh\cnd\vz)
\end{equation}
with $p(\vy\cnd\vh) = \mathcal{N}_\CC(\ma\vh,\msig)$ and $p_\vtheta(\vh\cnd\vz)$ as in~\eqref{eq:cg-vae}. The dependency of $p(\vy\cnd\vh,\vz)$ from $\vz$ can be dropped due to the Bayesian network in Fig.~\ref{fig:prob_graph} showing $\vy$ is conditionally independent of $\vz$ given $\vh$.
Since the multiplication of two Gaussian distributions is again Gaussian, $p_\vtheta(\vy,\vh\cnd\vz)=$
\begin{equation}
    \mathcal{N}_\CC \left( 
    \begin{bmatrix} \ma\vmu_\vtheta \\ \vmu_\vtheta \end{bmatrix},
    \begin{bmatrix}
        \msig^{-1} & -\msig^{-1}\ma \\
        -\ma\herm\msig^{-1} & \ma\herm\msig^{-1}\ma + \mc^{-1}_\vtheta
    \end{bmatrix}^{-1}
    \right)
\end{equation}
after some algebraic reformulations, which shows that $\vy$ and $\vh$ are jointly Gaussian given $\vz$.
We omit the $\vz$-argument here for notational brevity.
For the derivation of $\E_\vtheta[\vh\cnd\vz,\vy]$, we are interested in the conditional $p_\vtheta(\vh\cnd\vz,\vy)$.
Using standard results for jointly Gaussian distributions, the conditional is again Gaussian with the mean vector $\E_\vtheta[\vh\cnd\vz,\vy]=$
\begin{equation}
    \label{eq:app-lmmse-init}
    \vmu_\vtheta + \left( \ma\herm\msig^{-1}\ma + \mc^{-1}_\vtheta \right)^{-1} \ma\herm\msig^{-1} (\vy - \ma\vmu_\vtheta)
\end{equation}
and covariance matrix $(\ma\herm\msig^{-1}\ma + \mc^{-1}_\vtheta)^{-1}$.
Application of the matrix inversion lemma to~\eqref{eq:app-lmmse-init} and further algebraic reformulations yield~\eqref{eq:lmmse-vae}, concluding the derivation.

\subsection{Proof of Theorem \ref{cor:cme_bound}}
\label{app:cme_bound2}

\begin{proof}

Let us define the variables
\begin{align}
\B \Psi &= \ftt - \fmt,~~~\B \psi = \fto - \fmo,
\label{eq:variables_bound}
\\
\B \Gamma &= \fmt + \varsigma^2 \eye,
\label{eq:variables_bound2}
\end{align}
and denote $\Et[\cdot] := \E_{p_{\btheta}(\B z|\B y)}[\cdot]$ for notational convenience.
First, let us rewrite $\B T(\B z) = \B C_{\btheta}(\B z) (\B C_{\btheta}(\B z) + \varsigma^2\eye)\inv $ as
\begin{align}
    \B T(\B z)
    &= (\B C_{\btheta}(\B z) + \varsigma^2\eye)\inv \B C_{\btheta}(\B z)
    \label{eq:push_through}
    \\
    &= (\B \Gamma + \B \Psi)\inv \ftt 
    \\
    &= \left(\B \Gamma\inv - \B \Gamma\inv \B \Psi(\B \Gamma + \B \Psi)\inv  \right) \ftt 
    \label{eq:matrix_inversion_lemma}
    \\
    &= \B \Gamma\inv \fmt + \B \Gamma\inv \B \Psi \left(\eye - (\B \Gamma + \B \Psi)\inv \ftt\right)
\end{align}
by using the push-through identity in \eqref{eq:push_through} and the matrix inversion lemma in \eqref{eq:matrix_inversion_lemma}.
Using this result, we rewrite the \ac{cme} $\E[\B h \cnd \B y]$ from \eqref{eq:total_exp} in terms of the \ac{map}-\ac{vae} estimator $\hat{\B h}_{\text{VAE}}(\B y)$ from \eqref{eq:vae-estimator-mu} and an additive  error term as
\begin{align}
    &\E[\B h \cnd \B y] = \Et[\fto + \B T(\B z)(\B y - \fto)]
    \\
    &\begin{aligned}
        =&\underbrace{\fmo + \B \Gamma\inv \fmt (\B y - \fmo)}_{=\hat{\B h}_{\text{VAE}}(\B y)}
        \\ 
        &+ \Et[(\eye - \B \Gamma\inv \fmt) \B \psi] 
        \\
        &+ \Et[\B \Gamma\inv \B \Psi(\eye - (\B \Gamma + \B \Psi)\inv \ftt)(\B y - \fto)].
        \label{eq:cme_in_vae}
    \end{aligned}
\end{align}
We further note that we can simplify the term 
\begin{align}
    \eye - (\B \Gamma + \B \Psi)\inv \ftt)(\B y - \fto) &= \B y - \E[\B h | \B y, \B z]
    \\
    &= \E[\B n | \B y, \B z].
    \label{eq:bound_exp_sigma}
\end{align}
Thus, we get an upper bound on the expected Euclidean distance between the \mapvae estimator and the \ac{cme}  $\varepsilon = \E[\|\E[\B h \cnd \B y] - \hat{\B h}_{\text{VAE}}(\B y)\|_2]$ as
\begin{align}
    &\begin{aligned}
        \varepsilon &\leq \E\left[\|\eye - \B \Gamma\inv \fmt \|_2 \Et[\|\B \psi\|_2]\right]
        \\
        &~~~+ \E\left[\|\B \Gamma\inv\|_2 \cdot \|\Et[\B \Psi \E[\B n |\B y, \B z]]\|_2\right]
    \end{aligned}
    \\
    &\phantom{\bar{\varepsilon}}\leq \E\left[\tfrac{\varsigma^2}{\xi_{\text{min}} + \varsigma^2} \Et[\|\B \psi\|_2]\right]
    + \tfrac{1}{\varsigma^2}\E\left[ \Et[\|\B \Psi \E[\B n |\B y, \B z]\|_2]\right]
    \label{eq:summand_cor}
\end{align}
where we used the reformulation from \eqref{eq:bound_exp_sigma} and the bounds on the spectral norms 
\begin{align}
    \|\eye - \B \Gamma\inv \fmt \|_2 = \frac{\varsigma^2}{\xi_{\text{min}} + \varsigma^2},
    ~~~\|\B \Gamma\inv\|_2 \leq \frac{1}{\varsigma^2},
    \label{eq:bound_spectral_norm}
\end{align}
together with the triangle and Cauchy-Schwarz inequalities. Note that $\xi_{\text{min}}$ is a function of $\B y$ and the outer expectations are with respect to $p(\vy)$ if not denoted otherwise.
Thus, we employ H\"older's inequality for both summands in \eqref{eq:summand_cor} to get
\begin{align}
    &\begin{aligned}
        \varepsilon &\leq C_1\sqrt{\E[\Et[\|\B \psi\|_2^2]]} 
        \\
        &~~~+ \frac{1}{\varsigma^2} \sqrt{\E[\Et[\|\E[\B n | \B y, \B z]\|_2^2]]} \sqrt{\E[\Et[\|\B \Psi\|_2^2]]} 
    \end{aligned}
    \\
    &\phantom{\varepsilon} \leq C_1 \sqrt{\E[\Et[\|\B \psi\|_2^2]]} + \frac{1}{\varsigma^2} \sqrt{\E[\|\B n\|_2^2]} \sqrt{\E[\Et[\|\B \Psi\|_2^2]]} 
    \label{eq:bound_noise_vec}
    \\
     &\phantom{\varepsilon} = C_1 \sqrt{\E[\Et[\|\B \psi\|_2^2]]} + C_2 \sqrt{\E[\Et[\|\B \Psi\|_2^2]]}
\end{align}
where we used Jensen's inequality in combination with the law of total expectation in \eqref{eq:bound_noise_vec}, and with $C_1, C_2$ in \eqref{eq:c1_c2_exp}. 
By resubstituting the variables in \eqref{eq:variables_bound}, we get
\begin{align}
    &\begin{aligned}\label{eq:cond-bias-variance}
        \varepsilon \leq &C_1 \sqrt{\E[\Et[\|\fto - \fmo\|_2] ]}
        \\ & +C_2 \sqrt{\E[\Et[\|\ftt - \fmt\|_2^2]]}
    \end{aligned}
    \\
    &\phantom{\varepsilon}
    \leq (C_1L_1 + C_2L_2) \sqrt{\E[\Et[\|\B z - \B \mu_{\bphi}\|_2^2]]}
    \label{eq:cme_bound2}
\end{align}
by using the Lipschitz continuity \eqref{eq:lipschitz}.
By defining the first and second moments of the posterior distribution $p_{\btheta}(\B z \cnd \B y)$ as $\bar{\B \mu} = \Et[\B z]$ and $\bar{\B C} = \Et[(\B z - \bar{\B \mu} ) (\B z - \bar{\B \mu} )\herm]$, we write
\begin{align}
\label{eq:bound_end1}
    \sqrt{\E[\Et[\|\B z - \B\mu_{\bphi}\|_2^2]]} &= \sqrt{\E[\Et[\| \B z - \bar{\B \mu} + \bar{\B \mu} - \B\mu_{\bphi} \|_2^2]]}
    \\
    &\hspace{-1cm}\leq \sqrt{\E[\Et[\left(\|\B z - \bar{\B \mu} \|_2 + \| \bar{\B \mu} - \B\mu_{\bphi}\|_2\right)^2]]}
    \label{eq:first_summ}
    \\
    &\hspace{-1cm}\leq \sqrt{\tr(\bar{\B C})} + \sqrt{\E[\| \bar{\B \mu} - \B\mu_{\bphi}\|_2^2]}
    \label{eq:bound_end2}
\end{align}
since only one summand in \eqref{eq:first_summ} depends on $\B y$ or $\B z$, respectively. Plugging \eqref{eq:bound_end2} in \eqref{eq:cme_bound2} yields \eqref{eq:cme_bound_exp}, completing the~proof.

\end{proof}
\balance
\bibliographystyle{IEEEtran}
\bibliography{main}

\begin{thebibliography}{10}
\providecommand{\url}[1]{#1}
\csname url@samestyle\endcsname
\providecommand{\newblock}{\relax}
\providecommand{\bibinfo}[2]{#2}
\providecommand{\BIBentrySTDinterwordspacing}{\spaceskip=0pt\relax}
\providecommand{\BIBentryALTinterwordstretchfactor}{4}
\providecommand{\BIBentryALTinterwordspacing}{\spaceskip=\fontdimen2\font plus
\BIBentryALTinterwordstretchfactor\fontdimen3\font minus
  \fontdimen4\font\relax}
\providecommand{\BIBforeignlanguage}[2]{{%
\expandafter\ifx\csname l@#1\endcsname\relax
\typeout{** WARNING: IEEEtran.bst: No hyphenation pattern has been}%
\typeout{** loaded for the language `#1'. Using the pattern for}%
\typeout{** the default language instead.}%
\else
\language=\csname l@#1\endcsname
\fi
#2}}
\providecommand{\BIBdecl}{\relax}
\BIBdecl

\bibitem{Ruthotto2021}
L.~Ruthotto and E.~Haber, ``An introduction to deep generative modeling,''
  \emph{GAMM-Mitteilungen}, vol.~44, no.~2, p. e202100008, 2021.

\bibitem{Bishop2006}
C.~M. Bishop, \emph{{Pattern Recognition and Machine Learning}}.\hskip 1em plus
  0.5em minus 0.4em\relax Springer, 2006.

\bibitem{Rezende2014}
D.~J. Rezende, S.~Mohamed, and D.~Wierstra, ``{Stochastic Backpropagation and
  Approximate Inference in Deep Generative Models},'' in \emph{Proc. 31st Int.
  Conf. Mach. Learn.}, 2014.

\bibitem{Kingma2014}
D.~P. Kingma and M.~Welling, ``{Auto-Encoding Variational Bayes},'' in
  \emph{Proc. 2nd Int. Conf. Learn. Represent.}, 2014.

\bibitem{Goodfellow2020}
I.~Goodfellow \emph{et~al.}, ``{Generative Adversarial Networks},''
  \emph{Commun. ACM}, vol.~63, no.~11, pp. 139--144, 2020.

\bibitem{Song2020}
Y.~Song and S.~Ermon, ``Generative modeling by estimating gradients of the data
  distribution,'' in \emph{Adv. Neural Inf. Process. Syst.}, vol.~32, 2019.

\bibitem{Ongie2020}
G.~Ongie \emph{et~al.}, ``{Deep Learning Techniques for Inverse Problems in
  Imaging},'' \emph{IEEE J. Sel. Areas Inf. Theory}, vol.~1, no.~1, pp. 39--56,
  2020.

\bibitem{Bora2017}
A.~Bora, A.~Jalal, E.~Price, and A.~G. Dimakis, ``{Compressed Sensing using
  Generative Models},'' \emph{Proc. 34th Int. Conf. Mach. Learn.}, vol.~70, pp.
  537--546, 2017.

\bibitem{Jalal2021}
A.~Jalal \emph{et~al.}, ``{Robust Compressed Sensing MRI with Deep Generative
  Priors},'' in \emph{Adv. Neural Inf. Process. Syst.}, 2021.

\bibitem{Hand2018}
P.~Hand, O.~Leong, and V.~Voroninski, ``{Phase Retrieval Under a Generative
  Prior},'' in \emph{Adv. Neural Inf. Process. Syst.}, 2018.

\bibitem{Asim2020}
M.~Asim, F.~Shamshad, and A.~Ahmed, ``{Blind Image Deconvolution Using Deep
  Generative Priors},'' \emph{IEEE Trans. Comput. Imaging}, vol.~6, pp.
  1493--1506, 2020.

\bibitem{Koller2022}
M.~Koller, B.~Fesl, N.~Turan, and W.~Utschick, ``{An Asymptotically MSE-Optimal
  Estimator Based on Gaussian Mixture Models},'' \emph{IEEE Trans. Signal
  Process.}, vol.~70, pp. 4109--4123, 2022.

\bibitem{Balevi2021}
E.~Balevi and J.~G. Andrews, ``{Wideband Channel Estimation With a Generative
  Adversarial Network},'' \emph{IEEE Trans. Wirel. Commun.}, vol.~20, no.~5,
  pp. 3049--3060, 2021.

\bibitem{Balevi2021a}
E.~Balevi, A.~Doshi, A.~Jalal, A.~Dimakis, and J.~G. Andrews, ``{High
  Dimensional Channel Estimation Using Deep Generative Networks},'' \emph{IEEE
  J. Sel. Areas Commun.}, vol.~39, no.~1, pp. 18--30, 2021.

\bibitem{Doshi2022}
A.~S. Doshi, M.~Gupta, and J.~G. Andrews, ``{Over-the-Air Design of GAN
  Training for mmWave MIMO Channel Estimation},'' \emph{IEEE J. Sel. Areas Inf.
  Theory}, vol.~3, no.~3, pp. 557--573, 2022.

\bibitem{Arvinte2023}
M.~Arvinte and J.~I. Tamir, ``{MIMO Channel Estimation Using Score-Based
  Generative Models},'' \emph{IEEE Trans. Wirel. Commun.}, vol.~22, no.~6, pp.
  3698--3713, 2023.

\bibitem{Diskin2023}
T.~Diskin, Y.~C. Eldar, and A.~Wiesel, ``{Learning to Estimate Without Bias},''
  \emph{IEEE Trans. Signal Process.}, vol.~71, pp. 2162--2171, 2023.

\bibitem{Kay1993}
S.~M. Kay, \emph{{Fundamentals of Statistical Signal Processing: Estimation
  Theory}}.\hskip 1em plus 0.5em minus 0.4em\relax Englewood Cliffs, NJ:
  Prentice-Hall, Inc., 1993.

\bibitem{Zhao2017a}
S.~Zhao, J.~Song, and S.~Ermon, ``{Towards a Deeper Understanding of
  Variational Autoencoding Models},'' \emph{arXiv preprint arXiv:1702.08658},
  2017.

\bibitem{Baur2022}
M.~Baur, B.~Fesl, M.~Koller, and W.~Utschick, ``{Variational Autoencoder
  Leveraged MMSE Channel Estimation},'' in \emph{56th Asilomar Conf. Signals,
  Syst., Comput.}, 2022, pp. 527--532.

\bibitem{Rani2018}
M.~Rani, S.~B. Dhok, and R.~B. Deshmukh, ``{A Systematic Review of Compressive
  Sensing: Concepts, Implementations and Applications},'' \emph{IEEE Access},
  vol.~6, pp. 4875--4894, 2018.

\bibitem{1459065}
A.~Banerjee, X.~Guo, and H.~Wang, ``{On the Optimality of Conditional
  Expectation as a {Bregman} Predictor},'' \emph{IEEE Trans. Inf. Theory},
  vol.~51, no.~7, pp. 2664--2669, 2005.

\bibitem{Kingma2019}
D.~P. Kingma and M.~Welling, ``{An Introduction to Variational Autoencoders},''
  \emph{Found. Trends{\textregistered} Mach. Learn.}, vol.~12, no.~4, pp.
  307--392, 2019.

\bibitem{Bertsekas2014}
D.~P. Bertsekas and J.~N. Tsitsiklis, \emph{{Introduction to Probability}},
  2nd~ed.\hskip 1em plus 0.5em minus 0.4em\relax Nashua, NH: Athena Scientific,
  2008.

\bibitem{Yang2015}
J.~Yang \emph{et~al.}, ``{Compressive Sensing by Learning a Gaussian Mixture
  Model From Measurements},'' \emph{IEEE Trans. Image Process.}, vol.~24,
  no.~1, pp. 106--119, 2015.

\bibitem{Loeve1977}
M.~Lo{\`{e}}ve, \emph{{Probability Theory I}}, 4th~ed.\hskip 1em plus 0.5em
  minus 0.4em\relax Springer New York, NY, 1977.

\bibitem{Fuhrmann1991}
D.~Fuhrmann, ``{Application of Toeplitz covariance estimation to adaptive
  beamforming and detection},'' \emph{IEEE Trans. Signal Process.}, vol.~39,
  no.~10, pp. 2194--2198, 1991.

\bibitem{Ephraim1989}
Y.~Ephraim, D.~Malah, and B.-H. Juang, ``{On the application of hidden Markov
  models for enhancing noisy speech},'' \emph{IEEE Trans. Acoust.}, vol.~37,
  no.~12, pp. 1846--1856, 1989.

\bibitem{Neumann2018}
D.~Neumann, T.~Wiese, and W.~Utschick, ``{Learning The MMSE Channel
  Estimator},'' \emph{IEEE Trans. Signal Process.}, vol.~66, no.~11, pp.
  2905--2917, 2018.

\bibitem{Heckens2020}
A.~J. Heckens, S.~M. Krause, and T.~Guhr, ``{Uncovering the dynamics of
  correlation structures relative to the collective market motion},'' \emph{J.
  Stat. Mech. Theory Exp.}, vol. 2020, no.~10, p. 103402, 2020.

\bibitem{Fesl2022}
B.~Fesl \emph{et~al.}, ``{Channel Estimation based on Gaussian Mixture Models
  with Structured Covariances},'' in \emph{2022 56th Asilomar Conf. Signals,
  Syst. Comput.}, 2022, pp. 533--537.

\bibitem{Baur2023}
M.~Baur, N.~Turan, B.~Fesl, and W.~Utschick, ``{Channel Estimation in
  Underdetermined Systems Utilizing Variational Autoencoders},'' in \emph{2024
  IEEE Int. Conf. Acoust. Speech Signal Process.}, 2024, pp. 9031--9035.

\bibitem{Baur2023meas}
M.~Baur, B.~B{\"{o}}ck, N.~Turan, and W.~Utschick, ``{Variational Autoencoder
  for Channel Estimation: Real-World Measurement Insights},'' in \emph{2024
  27th Int. Work. Smart Antennas}.\hskip 1em plus 0.5em minus 0.4em\relax IEEE,
  2024, pp. 117--122.

\bibitem{Gr06}
R.~M. Gray, ``{Toeplitz and Circulant Matrices: {A} Review},'' \emph{Found. and
  Trends\textsuperscript{\textregistered} in Commun. and Inf. Theory}, no.~3,
  pp. 155--239, 2006.

\bibitem{Bock2024}
B.~B{\"{o}}ck, M.~Baur, N.~Turan, D.~Semmler, and W.~Utschick, ``{A Statistical
  Characterization of Wireless Channels Conditioned on Side Information},''
  \emph{arXiv preprint arXiv:2406.04282}, 2024.

\bibitem{Yin2016}
X.~Yin and C.~Xiang, \emph{{Propagation Channel Characterization, Parameter
  Estimation, and Modeling for Wireless Communications}}.\hskip 1em plus 0.5em
  minus 0.4em\relax Wiley, 2016.

\bibitem{Kermoal2002}
J.~Kermoal, L.~Schumacher, K.~Pedersen, P.~Mogensen, and F.~Frederiksen, ``{A
  Stochastic MIMO Radio Channel Model With Experimental Validation},''
  \emph{IEEE J. Sel. Areas Commun.}, vol.~20, no.~6, pp. 1211--1226, 2002.

\bibitem{Tse2005}
D.~Tse and P.~Viswanath, \emph{{Fundamentals of Wireless Communication}}.\hskip
  1em plus 0.5em minus 0.4em\relax New York, NY: Cambridge University Press,
  2005.

\bibitem{Bergstra2012}
J.~Bergstra and Y.~Bengio, ``{Random Search for Hyper-Parameter
  Optimization},'' \emph{J. Mach. Learn. Res.}, vol.~13, no.~10, pp. 281--305,
  2012.

\bibitem{Liaw2018}
R.~Liaw \emph{et~al.}, ``{Tune: A Research Platform for Distributed Model
  Selection and Training},'' \emph{arXiv preprint arXiv:1807.05118}, 2018.

\bibitem{Ioffe2015}
S.~Ioffe and C.~Szegedy, ``{Batch Normalization: Accelerating Deep Network
  Training by Reducing Internal Covariate Shift},'' in \emph{Proc. 32nd Int.
  Conf. Mach. Learn.}, 2015, pp. 448--456.

\bibitem{Salimans2016a}
T.~Salimans and D.~P. Kingma, ``{Weight Normalization: A Simple
  Reparameterization to Accelerate Training of Deep Neural Networks},''
  \emph{Adv. Neural Inf. Process. Syst.}, vol.~30, 2016.

\bibitem{Ba2016}
J.~L. Ba, J.~R. Kiros, and G.~E. Hinton, ``{Layer Normalization},'' in
  \emph{Adv. Neural Inf. Process. Syst. - Deep Learn. Symp.}, 2016.

\bibitem{Anwar2017}
S.~Anwar, K.~Hwang, and W.~Sung, ``{Structured Pruning of Deep Convolutional
  Neural Networks},'' \emph{ACM J. Emerg. Technol. Comput. Syst.}, vol.~13,
  no.~3, pp. 1--18, 2017.

\bibitem{3GPP2020}
3GPP, ``{Spatial channel model for Multiple Input Multiple Output (MIMO)
  simulations (Release 16)},'' 3rd Generation Partnership Project (3GPP), Tech.
  Rep. 25.996 V16.0.0, 2020.

\bibitem{Jaeckel2014}
S.~Jaeckel, L.~Raschkowski, K.~Borner, and L.~Thiele, ``{QuaDRiGa: A 3-D
  Multi-Cell Channel Model With Time Evolution for Enabling Virtual Field
  Trials},'' \emph{IEEE Trans. Antennas Propag.}, vol.~62, no.~6, pp.
  3242--3256, 2014.

\bibitem{QUADRIGA2016}
S.~Jaeckel \emph{et~al.}, ``{QuaDRiGa - Quasi Deterministic Radio Channel
  Generator, User Manual and Documentation},'' Fraunhofer Heinrich Hertz
  Institute, Tech. Rep. v2.6.1, 2021.

\bibitem{Busari2018}
S.~A. Busari, K.~M.~S. Huq, S.~Mumtaz, L.~Dai, and J.~Rodriguez,
  ``{Millimeter-Wave Massive MIMO Communication for Future Wireless Systems: A
  Survey},'' \emph{IEEE Commun. Surv. Tutorials}, vol.~20, no.~2, pp. 836--869,
  2018.

\bibitem{Donoho2010}
D.~L. Donoho, A.~Maleki, and A.~Montanari, ``{Message passing algorithms for
  compressed sensing: I. motivation and construction},'' in \emph{2010 IEEE
  Inf. Theory Work. Inf. Theory}.\hskip 1em plus 0.5em minus 0.4em\relax IEEE,
  2010, pp. 1--5.

\bibitem{Maleki2013}
A.~Maleki, L.~Anitori, Z.~Yang, and R.~G. Baraniuk, ``{Asymptotic Analysis of
  Complex LASSO via Complex Approximate Message Passing (CAMP)},'' \emph{IEEE
  Trans. Inf. Theory}, vol.~59, no.~7, pp. 4290--4308, 2013.

\bibitem{Fesl2021}
B.~Fesl, N.~Turan, M.~Koller, and W.~Utschick, ``{A Low-Complexity MIMO Channel
  Estimator with Implicit Structure of a Convolutional Neural Network},'' in
  \emph{22nd Int. Work. Signal Process. Adv. Wirel. Commun.}, 2021, pp. 11--15.

\bibitem{Fesl2024}
B.~Fesl, N.~Turan, B.~Bock, and W.~Utschick, ``{Channel Estimation for
  Quantized Systems based on Conditionally Gaussian Latent Models},''
  \emph{IEEE Trans. Signal Process.}, vol.~72, pp. 1475--1490, 2024.

\end{thebibliography}




\end{document}